\documentclass[12pt]{article} 
\usepackage{color}
\usepackage{amsfonts}

\newtheorem{definition}{Definition}[section] 
\newtheorem{example}{Example}[section] 
 
\newtheorem{lemma}{Lemma}[section] 
\newtheorem{theorem}[lemma]{Theorem} 
\newtheorem{corollary}[lemma]{Corollary} 
 
\newtheorem{algorithm}{Algorithm}[section]

\begin{document} 
 
\title{\Large \bf Termination Analysis of General Logic Programs for Moded Queries: A Dynamic Approach} 
 
\author{Yi-Dong Shen\\
{\small
Institute of Software, the Chinese Academy of Sciences, Beijing 100080, China}\\ 
{\small Email: ydshen@ios.ac.cn} \\[.06in] 
Danny De Schreye \\ 
{\small  Department of Computer Science, Celestijnenlaan 200 A, B-3001 Heverlee, Belgium}\\ 
{\small  Email: Danny.DeSchreye@cs.kuleuven.ac.be}
} 
 
\date{} 
 
\maketitle 
 
\begin{abstract}
The termination problem of a logic program can be addressed
in either a static or a dynamic way. A static approach performs termination analysis 
at compile time, while a dynamic approach characterizes and tests termination of a logic program
by applying a loop checking technique. In this paper, we present a novel
dynamic approach to termination analysis for general logic programs with moded queries.
We address several interesting questions, including how to formulate an SLDNF-derivation
for a moded query, how to characterize an infinite SLDNF-derivation with a moded query,
and how to apply a loop checking mechanism to cut infinite SLDNF-derivations for
the purpose of termination analysis. The proposed approach 
is very powerful and useful. It can be used 
(1) to test if a logic program terminates for a given concrete or moded query,
(2) to test if a logic program terminates for all concrete or moded queries, and
(3) to find all (most general) concrete/moded queries that are most likely  
terminating (or non-terminating).  

\bigskip

\end{abstract} 
 
\section{Introduction} 
\label{intr}
Given a logic program $P$, can we determine that $P$ terminates
for certain queries? This is the well-known termination problem in
logic programming. It is undecidable in general.
Two different ways have been explored in the literature to attack this problem. 
The first way is to perform termination analysis 
at compile time, thus referred to as a {\em static} approach \cite{DD93}, while
the other is to characterize and test termination of a logic program
by applying a loop checking technique, thus referred to as a 
{\em dynamic} approach \cite{shen-tocl}. Loop checking
is a technique for detecting and cutting infinite 
derivations at run time \cite{BAK91,shen001}.
Static termination analysis has been extensively studied in the literature
\cite{Apt1,Bezem92,BCF94,BS91,DV95,DDF93,DDV99,GC05,LS97,MN01,Pl90a,UVG88}
(see \cite{DD93} for a survey). However, although 
a number of loop checking mechanisms have been proposed 
\cite{BAK91,BDM92,MD96,Sa93,shen97,shen001,VG87,VL89},
it is only in \cite{shen-tocl} that the idea of using 
a loop checking technique for termination analysis 
is formally presented.

The intuition behind a dynamic approach is as follows.
Given a complete loop checking mechanism (that cuts any infinite 
derivations) and a set of queries, 
we run the program for each query while performing loop checking.
If the query evaluation terminates without cutting any derivations,
the program terminates for any query, 
otherwise it is potentially non-terminating for some queries.

In this paper, we are concerned with dynamic termination approaches.
The core of such an approach is a characterization of infinite
SLDNF-derivations, as any loop checking mechanism relies on it. 
In \cite{shen-tocl}, the first such characterization
is established for general logic programs. However, 
this characterization applies only
to concrete queries and cannot handle moded queries.
A moded query contains (abstract) atoms like $p({\cal I},T)$
where $T$ is a term (i.e., a constant, variable or function) and ${\cal I}$
is an input mode. An {\em input mode} stands for an arbitrary ground (i.e.\ variable-free)
term. Moded queries are commonly used in termination analysis of logic programs,
where to prove that a logic program terminates
for a moded query $p({\cal I},T)$ is to prove that 
the program terminates for any (concrete) query $p(t,T)$
where $t$ is a ground term. Consider the following logic program:
\begin{tabbing}
$\qquad$ \= $P_0:$ $\quad$ \= $p(a).$ \`$C_{p_1}$\\
\> \>         $p(f(X))\leftarrow p(X)$. \`$C_{p_2}$
\end{tabbing}
For any concrete query $p(t)$, 
evaluating $p(t)$ over $P_0$ will terminate. However, 
we cannot evaluate a moded query $p({\cal I})$ 
while applying a loop checking mechanism to infer 
that $P_0$ terminates for $p({\cal I})$.

In this paper, we present a dynamic approach
to characterizing and testing termination of logic programs 
for moded queries. For a logic program $P$ and
a moded query $Q_0$, the first issue we address is how to formulate an SLDNF-derivation
for $Q_0$. We will introduce a framework called a {\em moded-query forest},
which consists of all (generalized) SLDNF-trees 
rooted at a ground instance of $Q_0$. An SLDNF-derivation for $Q_0$ is defined over
the moded-query forest such that $P$
terminates for $Q_0$ if and only if the moded-query 
forest contains no infinite SLDNF-derivations.

A moded-query forest may have an infinite number of SLDNF-trees, so it is infeasible to test
termination of a logic program by traversing the moded-query forest.
We will introduce a compact approximation for a 
moded-query forest, called a {\em moded generalized SLDNF-tree}. The key idea is to 
treat an input mode as a special variable like a Skolem constant. 
As a result, top-down derivations for a moded query can be 
constructed in the same way as the ones for a concrete query.
A characterization of termination of a logic program for moded queries is then established
in terms of some novel properties of a moded generalized SLDNF-tree.

The paper is organized as follows. Section \ref{pre}
reviews some basic concepts including generalized SLDNF-trees.
Section \ref{sec2} establishes a characterization for logic programs 
with moded queries. Section \ref{sec4} develops an algorithm for testing
termination of logic programs for moded queries. Section \ref{related-work}
describes some closely related work, and Section \ref{conclusion} concludes.

\section{Preliminaries}
\label{pre}

We assume the reader is familiar with standard terminology of 
logic programs as described in \cite{Ld87}.
Variables begin with a capital letter $X,Y,Z,U,V$ or $I$, and predicate, function 
and constant symbols with a lower case letter. Let $A$ be an atom/term. 
The size of $A$, denoted $|A|$, is the 
number of occurrences of function symbols, variables and constants in $A$.
A list is of the form 
$[]$ or $[T|L]$ where $T$ is a term and $L$ is a list. 
For our purpose, the symbols $[$, $]$ and $|$ in a list  
are treated as function symbols.
Two atoms are called {\em variants} if they are the same
up to variable renaming.  
A (general) logic program $P$ is a finite set
of clauses of the form $A\leftarrow L_1,..., L_n$
where $A$ is an atom and $L_i$s are literals.
Throughout the paper, we consider only Herbrand models.
The Herbrand universe and Herbrand base of $P$ are denoted by
$HU(P)$ and $HB(P)$, respectively.
 
A goal $G_i$ is a headless clause
$\leftarrow L_1,..., L_n$ where each literal $L_j$ is called a subgoal.
The initial goal, $G_0$, is called
a top goal. Without loss of generality, we assume
that a top goal consists only of one atom. 
For a top goal $G_0 =\leftarrow A$, $Q_0 = A$ is called a query.
$Q_0$ is a {\em moded query} if some arguments of $A$ 
are input modes (in this case, $A$ is called an {\em abstract} atom); 
otherwise, it is a {\em concrete query}. 
An input mode always begins with a letter $\cal I$.  


Throughout the paper, we choose to use the best-known 
{\em depth-first, left-most} control strategy (used in Prolog)
to describe our approach (it can be adapted
to any other fixed control strategies).  
So the {\em selected subgoal} in each goal is the 
left-most subgoal.

A node in a top-down derivation tree (like SLDNF-trees) 
is represented by $N_i:G_i$ where $N_i$ is the name 
of the node and $G_i$ is a goal labeling the node. 
An ancestor-descendant relation is defined
on selected subgoals. $A$ is an ancestor subgoal of $B$,
denoted $A\prec_{anc}B$, if the proof of $A$ 
goes via the proof of $B$. The ancestor-descendant relation is expressed using 
an {\em ancestor list}. The ancestor list of a subgoal $B$ at a node $N$, 
denoted $AL_{B@N}$, consists of all pairs $(M, A)$
such that $A$ at $M$ is an ancestor subgoal of $B$ at $N$.

To characterize infinite derivations more precisely, in \cite{shen-tocl} 
standard SLDNF-trees \cite{Ld87} are extended to SLDNF$^*$-trees.
Informally, an SLDNF$^*$-tree is an SLDNF-tree except that
each node $N_i$ is associated with an ancestor list $AL_{L_j@N_i}$ 
for each subgoal $L_j$. In particular, let $L_1=\neg A$ be a selected subgoal at $N_i$, 
then a subsidiary child SLDNF$^*$-tree $T_{N_{i+1}:\leftarrow A}$ 
rooted at $N_{i+1}:\leftarrow A$ will be
built for solving this negative subgoal. 
Compared with a standard subsidiary SLDNF-tree $ST$ for $\neg A$, 
$T_{N_{i+1}:\leftarrow A}$ has two distinct features.
First, $N_{i+1}$ inherits the ancestor list $AL_{L_1@N_i}$. 
This mechanism bridges the ancestor-descendant
relationships across SLDNF$^*$-trees and is especially 
useful in identifying infinite derivations across SLDNF$^*$-trees.
Second, $T_{N_{i+1}:\leftarrow A}$ terminates at the first success leaf, 
so it may not include all branches of $ST$. 
This pruning mechanism (used in Prolog) is very
useful in not only improving 
the efficiency of query evaluation but also avoiding some possible 
infinite derivations (see Example \ref{eg-p2}). 
 
\begin{definition}[\cite{shen-tocl}]
{\em
Let $P$ be a logic program, $G_0$ a top goal, and
$T_{N_0:G_0}$ the SLDNF$^*$-tree for $P \cup \{G_0\}$.
A {\em generalized SLDNF-tree} for $P\cup \{G_0\}$, 
denoted $GT_{G_0}$, is rooted at $N_0:G_0$ and consists of $T_{N_0:G_0}$ 
along with all its descendant SLDNF$^*$-trees, where
parent and child SLDNF$^*$-trees are connected
via ``$\cdot\cdot\cdot\triangleright$". In $GT_{G_0}$ any path
starting at the root node $N_0:G_0$ (and ending at either 
a leaf or non-leaf node) is called a {\em generalized SLDNF-derivation}. 
} 
\end{definition}

``$\cdot\cdot\cdot\triangleright$" is called a {\em negation arc}.
For simplicity, in the sequel by a derivation we refer to a 
generalized SLDNF-derivation. 
Moreover, for any node $N_i:G_i$ we use 
$L_i^1$ to refer to the selected (i.e.\ the left-most) subgoal in $G_i$.
A derivation step is denoted by $N_i:G_i\Rightarrow_C N_{i+1}:G_{i+1}$,
meaning that applying clause $C$ to $L_i^1$ produces $N_{i+1}:G_{i+1}$.
For a substitution of two variables, $X$ in $L_i^1$ and $Y$ in (the
head of) $C$, we always use $X$ to substitute for $Y$, i.e. $Y/X$. 

\section{Characterizing Termination of Logic Programs for Moded Queries} 
\label{sec2}
In \cite{shen-tocl}, a characterization of termination 
of logic programs is established for concrete queries.
We reproduce the characterization and then extend it to the case of moded queries. 

\begin{definition}  
\label{symbol-seq}
{\em  
Let $T$ be a term or an atom and $S$ be  
a string that consists of all predicate symbols, function  
symbols, constants 
and variables in $T$, which is obtained 
by reading these symbols sequentially from left to  
right. The {\em symbol string} of $T$, denoted 
$S_T$, is the string $S$ with every variable  
replaced by ${\cal X}$. 
} 
\end{definition} 

For instance, let $T_1=a$, 
$T_2=f(X,g(X,f(a,Y)))$ and $T_3=[X,a]$. 
Then $S_{T_1}=a$, $S_{T_2}=f{\cal X}g{\cal X}fa{\cal X}$  
and $S_{T_3}=[{\cal X}|[a|[]]]$. Note that  
$[X,a]$ is a simplified representation for the 
list $[X|[a|[]]]$. 
 
\begin{definition}
\label{sub-seq} 
{\em
Let $S_{T_1}$ and $S_{T_2}$ be two symbol strings. 
$S_{T_1}$ is a {\em projection} of $S_{T_2}$, denoted  
$S_{T_1}\subseteq_{proj}S_{T_2}$,  
if $S_{T_1}$ is obtained from $S_{T_2}$ by
removing zero or more elements.
} 
\end{definition} 
 
\begin{definition}  
\label{gvar} 
{\em
Let $A_1=p(.)$ and $A_2=p(.)$  
be two atoms. $A_1$ is said to {\em loop into} $A_2$, 
denoted $A_1\leadsto_{loop}A_2$, if  
$S_{A_1}\subseteq_{proj}S_{A_2}$. 
Let $N_i:G_i$ and $N_j:G_j$ be two nodes in a  
derivation with $L_i^1\prec_{anc}L_j^1$ and 
$L_i^1 \leadsto_{loop} L_j^1$. 
Then $G_j$ is called a {\em loop goal} of $G_i$. 
}
\end{definition} 
 
Observe that if $A_1 \leadsto_{loop} A_2$ then $|A_1|\leq |A_2|$, and that 
if $G_3$ is a loop goal of $G_2$ that is a loop goal 
of $G_1$ then $G_3$ is a loop goal of $G_1$. 
Since a logic program has only
a finite number of clauses, an infinite derivation results 
from repeatedly applying the same set of clauses, which leads to either
infinite repetition of selected variant subgoals or  
infinite repetition of selected subgoals with recursive 
increase in term size. By recursive increase of term size 
of a subgoal $A$ from a subgoal $B$ we mean that $A$ is $B$ 
with a few function/constant/variable symbols added  
and possibly with some variables changed to 
different variables. Such crucial dynamic 
characteristics of an infinite derivation  
are captured by loop goals.

\begin{theorem}[\cite{shen-tocl}] 
\label{th1} 
Let $G_0=\leftarrow A$ be a top goal with $A$ a concrete query.
Any infinite derivation $D$ in $GT_{G_0}$ is of the form 
\begin{tabbing} 
$\qquad$ $N_0:G_0\Rightarrow_{C_0} ... N_{g_1}:G_{g_1}\Rightarrow_{C_1}   
...N_{g_2}:G_{g_2}\Rightarrow_{C_2} ... N_{g_3}:G_{g_3}\Rightarrow_{C_3} ...$ 
\end{tabbing} 
such that for any $j \geq 1$, $G_{g_{j+1}}$ is a loop goal of $G_{g_j}$. 
\end{theorem}

This theorem leads to the following immediate result.

\begin{corollary}[Characterization for a concrete query \cite{shen-tocl}] 
\label{th-iff} 
A logic program $P$ terminates for a concrete query $Q_0$ if and only if  
$GT_{G_0}$ has no infinite derivation of the form 
\begin{tabbing} 
$\qquad$ $N_0:G_0\Rightarrow_{C_0} ... N_{g_1}:G_{g_1}\Rightarrow_{C_1}   
...N_{g_2}:G_{g_2}\Rightarrow_{C_2} ... N_{g_3}:G_{g_3}\Rightarrow_{C_3} ...$ 
\end{tabbing} 
such that for any $j \geq 1$, $G_{g_{j+1}}$ is a loop goal of $G_{g_j}$. 
\end{corollary} 

Let $pred(P)$ be the set of predicate symbols in $P$
and let $CQ(P)$ contain a concrete query $p(X_1,...,X_n)$ for each $n$-nary predicate 
symbol $p$ in $pred(P)$. Note that $CQ(P)$ is finite, as $pred(P)$ is finite.
Since $CQ(P)$ covers all most general concrete queries for $P$,
it is immediate that $P$ terminates for any concrete queries
if and only if it terminates for all queries in $CQ(P)$. 

In order to extend Corollary \ref{th-iff} to handle moded queries,
we first define derivations for a moded query. 

\begin{definition} 
\label{mod-tree} 
{\em
Let $P$ be a logic program and $Q_0=p({\cal I}_1, ..., {\cal I}_m, T_1, ..., T_n)$
a moded query. The {\em moded-query forest} for $Q_0$ over $P$,
denoted $MF_{Q_0}$, consists of all generalized SLDNF-trees built from
$P\cup \{G_0\}$, where $G_0 = \leftarrow p(t_1, ..., t_m, T_1, ..., T_n)$
with all $t_i$s being ground terms from $HU(P)$. A derivation for $Q_0$
is a derivation in any generalized SLDNF-tree of $MF_{Q_0}$.  
}
\end{definition}

Therefore, a logic program $P$ terminates for a moded query $Q_0$ if and only if
$MF_{Q_0}$ has no infinite derivations.

\begin{example}
\label{eg0}
{\em
Consider the logic program $P_0$ given in Section \ref{intr}.
Let $p({\cal I})$ be a moded query. 
The moded-query forest $MF_{p({\cal I})}$ consists of generalized SLDNF-trees
$GT_{\leftarrow p(a)}$, $GT_{\leftarrow p(f(a))}$, etc., as shown in Figure \ref{fig0}
where for simplicity the symbol $\leftarrow$ in each goal 
and all ancestor lists attached to each node are omitted.
Note that $MF_{p({\cal I})}$ has an infinite number of generalized SLDNF-trees.
However, any individual tree, $GT_{G_0}$ with 
$G_0=\leftarrow p(\underbrace{f(f(...f}_{n \ items}(a)...)))$ ($n\geq 0$), is finite.
$MF_{p({\cal I})}$ contains no infinite derivations,
thus $P_0$ terminates for $p({\cal I})$. 
}
\end{example}
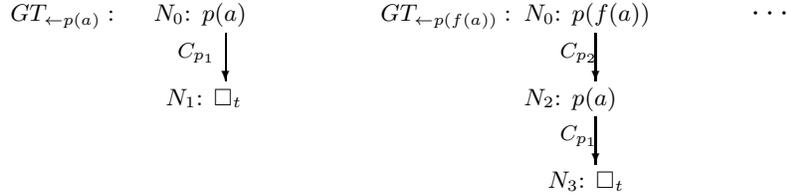
\begin{figure}[htb]
\begin{center}
\setlength{\unitlength}{3947sp}%
\begingroup\makeatletter\ifx\SetFigFont\undefined%
\gdef\SetFigFont#1#2#3#4#5{%
  \reset@font\fontsize{#1}{#2pt}%
  \fontfamily{#3}\fontseries{#4}\fontshape{#5}%
  \selectfont}%
\fi\endgroup%
\begin{picture}(4650,1215)(151,-511)
\thinlines
{\color[rgb]{0,0,0}\put(1501,464){\vector( 0,-1){300}}
}%
{\color[rgb]{0,0,0}\put(3826,464){\vector( 0,-1){300}}
}%
{\color[rgb]{0,0,0}\put(3826,-61){\vector( 0,-1){300}}
}%
\put(3601,314){\makebox(0,0)[lb]{\smash{\SetFigFont{8}{9.6}{\rmdefault}{\mddefault}{\updefault}{\color[rgb]{0,0,0}$C_{p_2}$}%
}}}
\put(3601,-211){\makebox(0,0)[lb]{\smash{\SetFigFont{8}{9.6}{\rmdefault}{\mddefault}{\updefault}{\color[rgb]{0,0,0}$C_{p_1}$}%
}}}
\put(3526,-511){\makebox(0,0)[lb]{\smash{\SetFigFont{9}{10.8}{\rmdefault}{\mddefault}{\updefault}{\color[rgb]{0,0,0}$N_3$:  $\Box_t$ }%
}}}
\put(3376, 14){\makebox(0,0)[lb]{\smash{\SetFigFont{9}{10.8}{\rmdefault}{\mddefault}{\updefault}{\color[rgb]{0,0,0}$N_2$:  $p(a)$ }%
}}}
\put(3376,539){\makebox(0,0)[lb]{\smash{\SetFigFont{9}{10.8}{\rmdefault}{\mddefault}{\updefault}{\color[rgb]{0,0,0}$N_0$:  $p(f(a))$ }%
}}}
\put(1126, 14){\makebox(0,0)[lb]{\smash{\SetFigFont{9}{10.8}{\rmdefault}{\mddefault}{\updefault}{\color[rgb]{0,0,0}$N_1$:  $\Box_t$ }%
}}}
\put(1051,539){\makebox(0,0)[lb]{\smash{\SetFigFont{9}{10.8}{\rmdefault}{\mddefault}{\updefault}{\color[rgb]{0,0,0}$N_0$:  $p(a)$ }%
}}}
\put(1201,314){\makebox(0,0)[lb]{\smash{\SetFigFont{8}{9.6}{\rmdefault}{\mddefault}{\updefault}{\color[rgb]{0,0,0}$C_{p_1}$}%
}}}
\put(4801,539){\makebox(0,0)[lb]{\smash{\SetFigFont{12}{14.4}{\rmdefault}{\mddefault}{\updefault}{\color[rgb]{0,0,0}$\cdots$ }%
}}}
\put(2476,539){\makebox(0,0)[lb]{\smash{\SetFigFont{9}{10.8}{\rmdefault}{\mddefault}{\updefault}{\color[rgb]{0,0,0}$GT_{\leftarrow p(f(a))}:$}%
}}}
\put(151,539){\makebox(0,0)[lb]{\smash{\SetFigFont{9}{10.8}{\rmdefault}{\mddefault}{\updefault}{\color[rgb]{0,0,0}$GT_{\leftarrow p(a)}:$}%
}}}
\end{picture}
\end{center}
\caption{The moded-query forest $MF_{p({\cal I})}$ for a moded query $p({\cal I})$.}\label{fig0} 
\end{figure} 

In a moded-query forest, all input modes are instantiated
into ground terms in $HU(P)$. When $HU(P)$ is infinite,
the moded-query forest would contain infinitely many
generalized SLDNF-trees. Thus it is infeasible to check
termination of a logic program for a moded query by applying
Corollary \ref{th-iff} over a moded-query forest.
An ideal way is to directly evaluate input modes
and build a compact generalized SLDNF-tree for a moded query.
Unfortunately, query evaluation in logic programming
accepts only terms as arguments of a top goal $- $ 
an input mode $\cal I$ is not directly evaluable.

Observe the following property of an input mode: 
it stands for an arbitrary ground term, that is,
it can be any term from $HU(P)$.
Therefore, during query evaluation it can be instantiated against any term.
This suggests that we may approximate the effect of an input mode 
by treating it as a special variable like a {\em Skolem constant}. 
A Skolem constant is an unknown constant and behaves like a variable.\footnote{The
two-faced feature of a Skolem constant is very useful. It is a special
constant, thus can appear in a negative subgoal without incurring floundering \cite{chan88}.
It is a special variable, thus can be instantiated against any term.}   
As a result, top-down derivations for a moded query can be 
constructed in the same way as the ones for a concrete query, 
where an input mode $\cal I$ is treated as a special variable $I$. 
 
\begin{definition} 
\label{mod-der} 
{\em
Let $P$ be a logic program and $Q_0=p({\cal I}_1, ..., {\cal I}_m, T_1, ..., T_n)$
a moded query. The {\em moded generalized SLDNF-tree} for
$Q_0$ over $P$ is defined to be the generalized SLDNF-tree $GT_{G_0}$ for
$P\cup \{G_0\}$, where $G_0 = \leftarrow p(I_1, ..., I_m, T_1, ..., T_n)$
with all $I_i$s being distinct variables not occurring in any $T_j$.
The variables $I_1, ..., I_m$ for the input modes
${\cal I}_1, ..., {\cal I}_m$ are called {\em input variables}.
}
\end{definition}

In a moded generalized SLDNF-tree, an input variable $I$ 
may be substituted by either a ground term $t$ or a non-ground function $f(.)$ 
(note that $I$ will never be substituted by a non-input variable).
If $I$ is substituted by $f(.)$, all variables in $f(.)$ are also called input 
variables. 

\begin{definition}
{\em
Let $D$ be a derivation in a moded generalized SLDNF-tree. A {\em moded instance}
of $D$ is a derivation obtained from $D$ by first 
instantiating all input variables at the root node 
with ground terms and then passing the instantiation down to the other input
variables along the derivation $D$.
}
\end{definition}

Let $Q_0 = p({\cal I}_1, ..., {\cal I}_m, T_1, ..., T_n)$ be a moded query. 
Any moded instance of a derivation $D$ for $Q_0$ is a derivation 
rooted at $N_0:p(t_1, ..., t_m, T_1, ..., T_n)$, where
all $t_i$s are ground terms from $HU(P)$. This means that 
any moded instance is a derivation in a moded-query forest $MF_{Q_0}$.

\begin{example}
\label{eg1}
{\em
Consider the logic program $P_0$ again.
Let $Q_0= p({\cal I})$ be a moded query. Then $G_0 = \leftarrow p(I)$.
The moded generalized SLDNF-tree $GT_{G_0}$ is depicted in Figure \ref{fig1}.
Since $I$ is an input variable, $X_2$ is an input variable, too
(due to the mgu (most general unifier)  $\theta_2$). For the same reason, 
all $X_{2i}$s are input variables ($i>0$). $GT_{G_0}$ has 
the following infinite derivation:  
\[N_0:p(I)\Rightarrow_{C_{p_2}} N_2:p(X_2) \Rightarrow_{C_{p_2}} N_4:p(X_4) \Rightarrow_{C_{p_2}} \cdots\]    
By instantiating $I$ with different ground terms,
we obtain different moded instances from this derivation.
For example, instantiating $I$ with $a$, $f(a)$ and $f(f(a))$ 
respectively yields the following moded instances:
\begin{tabbing} 
$\qquad\quad$ \= $N_0:p(a)$.\\
\> $N_0:p(f(a))\Rightarrow_{C_{p_2}} N_2:p(a)$. \\ 
\> $N_0:p(f(f(a)))\Rightarrow_{C_{p_2}} N_2:p(f(a)) \Rightarrow_{C_{p_2}} N_4:p(a)$.
\end{tabbing}
All these moded instances are derivations 
in the moded-query forest $MF_{Q_0}$ of Figure~\ref{fig0}. 
}
\end{example}
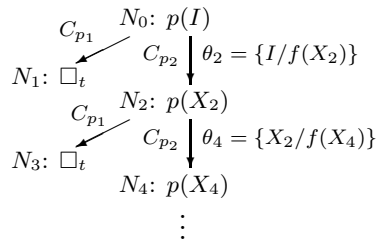
\begin{figure}[htb]
\begin{center}
\setlength{\unitlength}{3947sp}%
\begingroup\makeatletter\ifx\SetFigFont\undefined%
\gdef\SetFigFont#1#2#3#4#5{%
  \reset@font\fontsize{#1}{#2pt}%
  \fontfamily{#3}\fontseries{#4}\fontshape{#5}%
  \selectfont}%
\fi\endgroup%
\begin{picture}(1200,1446)(2176,-811)
\thinlines
{\color[rgb]{0,0,0}\put(3301,464){\vector( 0,-1){300}}
}%
{\color[rgb]{0,0,0}\put(3301,-61){\vector( 0,-1){300}}
}%
{\color[rgb]{0,0,0}\put(2921,-61){\vector(-2,-1){336}}
}%
{\color[rgb]{0,0,0}\put(2921,464){\vector(-2,-1){336}}
}%
\put(3226,-811){\makebox(0,0)[lb]{\smash{\SetFigFont{12}{14.4}{\rmdefault}{\mddefault}{\updefault}{\color[rgb]{0,0,0}$\vdots$ }%
}}}
\put(2551,-61){\makebox(0,0)[lb]{\smash{\SetFigFont{8}{9.6}{\rmdefault}{\mddefault}{\updefault}{\color[rgb]{0,0,0}$C_{p_1}$}%
}}}
\put(2476,464){\makebox(0,0)[lb]{\smash{\SetFigFont{8}{9.6}{\rmdefault}{\mddefault}{\updefault}{\color[rgb]{0,0,0}$C_{p_1}$}%
}}}
\put(2176,164){\makebox(0,0)[lb]{\smash{\SetFigFont{9}{10.8}{\rmdefault}{\mddefault}{\updefault}{\color[rgb]{0,0,0}$N_1$:  $\Box_t$ }%
}}}
\put(2176,-361){\makebox(0,0)[lb]{\smash{\SetFigFont{9}{10.8}{\rmdefault}{\mddefault}{\updefault}{\color[rgb]{0,0,0}$N_3$:  $\Box_t$ }%
}}}
\put(3376,-211){\makebox(0,0)[lb]{\smash{\SetFigFont{8}{9.6}{\rmdefault}{\mddefault}{\updefault}{\color[rgb]{0,0,0}$\theta_4 = \{X_2/f(X_4)\}$}%
}}}
\put(2851,-511){\makebox(0,0)[lb]{\smash{\SetFigFont{9}{10.8}{\rmdefault}{\mddefault}{\updefault}{\color[rgb]{0,0,0}$N_4$:  $p(X_4)$ }%
}}}
\put(3001,-211){\makebox(0,0)[lb]{\smash{\SetFigFont{8}{9.6}{\rmdefault}{\mddefault}{\updefault}{\color[rgb]{0,0,0}$C_{p_2}$}%
}}}
\put(3376,314){\makebox(0,0)[lb]{\smash{\SetFigFont{8}{9.6}{\rmdefault}{\mddefault}{\updefault}{\color[rgb]{0,0,0}$\theta_2 = \{I/f(X_2)\}$}%
}}}
\put(3001,314){\makebox(0,0)[lb]{\smash{\SetFigFont{8}{9.6}{\rmdefault}{\mddefault}{\updefault}{\color[rgb]{0,0,0}$C_{p_2}$}%
}}}
\put(2851, 14){\makebox(0,0)[lb]{\smash{\SetFigFont{9}{10.8}{\rmdefault}{\mddefault}{\updefault}{\color[rgb]{0,0,0}$N_2$:  $p(X_2)$ }%
}}}
\put(2851,539){\makebox(0,0)[lb]{\smash{\SetFigFont{9}{10.8}{\rmdefault}{\mddefault}{\updefault}{\color[rgb]{0,0,0}$N_0$:  $p(I)$}%
}}}
\end{picture}
\end{center}
\caption{The moded generalized SLDNF-tree $GT_{\leftarrow p(I)}$ 
for a moded query $p({\cal I})$.}\label{fig1} 
\end{figure} 

In a moded generalized SLDNF-tree $GT_{G_0}$ as shown in Figure \ref{fig1}, 
a moded query $p({\cal I})$ is approximated by a concrete query $p(I)$. 
Since $p(I)$ is more general than $p({\cal I})$ in the sense that 
$p({\cal I})$ covers only all ground instances of $p(I)$, $GT_{G_0}$ 
may contain some more general derivations not covered by $MF_{Q_0}$.   
So we have the following immediate result.

\begin{theorem}
\label{th2}
Let $MF_{Q_0}$ and $GT_{G_0}$ be the moded-query forest and the
moded generalized SLDNF-tree for $Q_0$ over $P$, respectively.
If $MF_{Q_0}$ has an infinite derivation $D'$,
$GT_{G_0}$ has an infinite derivation $D$ with 
$D'$ as a moded instance. But conversely, 
it is not necessarily true that if
$GT_{G_0}$ has an infinite derivation then
$MF_{Q_0}$ has an infinite derivation. 
\end{theorem}

Our goal is to establish a characterization of 
infinite derivations for a moded query such that 
the converse part of Theorem \ref{th2} is true under some conditions.

Consider the infinite derivation in Figure \ref{fig1} again. The input variable $I$
is substituted by $f(X_2)$, $X_2$ is then substituted by $f(X_4)$, \ldots. 
The substitutions go recursively and produce an infinite chain of substitutions for $I$ 
of the form $I/f(X_2), X_2/f(X_4),$ \ldots. The following lemma shows
that infinite derivations containing such an infinite chain of substitutions 
have no infinite moded instances.
 
\begin{lemma} 
\label{lem1}
If a derivation $D$ in a moded generalized SLDNF-tree 
$GT_{G_0}$ is infinite but none of its moded instances
is infinite, then there is an input variable $I$ such that $D$ contains
an infinite chain of substitutions for $I$ of the form
\begin{equation}
\label{ins-chain}
I/f_1(...,Y_1,...), ..., Y_1/f_2(...,Y_2,...), ..., Y_{i-1}/f_i(...,Y_i,...), ...
\end{equation}
(some $f_i$s would be the same). 
\end{lemma}

\noindent {\bf Proof:}
We distinguish four types of substitution chains for an input variable $I$ in $D$:
\begin{enumerate}
\item
\label{sub1}
$X_1/I, ..., X_m/I$ or $X_1/I, ..., X_i/I,$ \ldots. That is, $I$ is
never substituted by any terms.

\item
\label{sub2}
$X_1/I, ..., X_m/I, I/t$ where $t$ is a ground term. That is, $I$ is substituted by
a ground term.

\item
\label{sub3}
$X_1/I, ...,X_m/I, I/f_1(...,Y_1,...), ..., Y_1/f_2(...,Y_2,...), ..., Y_{n-1}/f_n(...,Y_n,...)$, ..., where
$f_n(...,Y_n,...)$ is the last non-ground function in the substitution chain for $I$ in $D$.
In this case, $I$ is recursively substituted by a finite number of functions.

\item
\label{sub4}
$X_1/I, ..., X_m/I, I/f_1(...,Y_1,...), ..., Y_1/f_2(...,Y_2,...), ..., Y_{i-1}/f_i(...,Y_i,...),$ \ldots.
In this case, $I$ is recursively substituted by an infinite number of functions. 
\end{enumerate}
For type \ref{sub1}, $D$ retains its infinite extension for whatever ground term we replace $I$ with.
For type \ref{sub2}, $D$ retains its infinite extension when we use $t$ to replace $I$.
To sum up, for any input variable $I$ whose substitution chain is of 
type \ref{sub1} or of type \ref{sub2}, there is a ground term $t$ such that 
replacing $I$ with $t$ does not affect the infinite extension of $D$.
In this case, replacing $I$ in $D$ with $t$ leads to an infinite derivation less general than $D$. 
 
For type \ref{sub3}, note that all variables appearing in the $f_i(.)$s are
input variables. Since $f_n(...,Y_n,...)$ is the last non-ground 
function in the substitution chain for $I$ in $D$,
the substitution chain for every variable $Y_n$ in $f_n(...,Y_n,...)$ 
is either of type \ref{sub1} or of type \ref{sub2}.
Therefore, we can replace each $Y_n$
with an appropriate ground term $t_n$ without affecting the infinite extension of $D$.
After this replacement, $D$ becomes $D_n$ and $f_n(...,Y_n,...)$ 
becomes a ground term $f_n(...,t_n,...)$.
Now $f_{n-1}(...,Y_{n-1},...)$ is the last non-ground
function in the substitution chain for $I$ in $D_n$.
Repeating the above replacement recursively, we will obtain an infinite derivation
$D_1$, which is $D$ with all variables in the $f_i(.)$s replaced with a ground term.
Assume $f_1(...,Y_1,...)$ becomes a ground term $t$ in $D_1$.
Then the substitution chain for $I$ in $D_1$ is of type \ref{sub2}. 
So replacing $I$ with $t$ in $D_1$ leads to an infinite derivation $D_0$.

The above constructive proof shows that if the substitution chains for all input variables in $D$
are of type \ref{sub1}, \ref{sub2} or \ref{sub3}, then $D$ must have an infinite moded instance.
Since $D$ has no infinite moded instance, there must exist an input variable $I$ 
whose substitution chain in $D$ is of type \ref{sub4}. That is,
$I$ is recursively substituted by an infinite number of functions.
Note that some $f_i$s would be the same because a logic program has only a finite number
of function symbols. This concludes the proof.
$\Box$

\bigskip

We are ready to introduce the following principal result. 

\begin{theorem} 
\label{th-main} 
Let $MF_{Q_0}$ and $GT_{G_0}$ be the moded-query forest and the
moded generalized SLDNF-tree for $Q_0$ over $P$, respectively.
$MF_{Q_0}$ has an infinite derivation 
if and only if $GT_{G_0}$ has an infinite derivation $D$ of the form 
\begin{equation} 
\label{eq2}
N_0:G_0\Rightarrow_{C_0} ... N_{g_1}:G_{g_1}\Rightarrow_{C_1}    
...N_{g_2}:G_{g_2}\Rightarrow_{C_2} ... N_{g_3}:G_{g_3}\Rightarrow_{C_3} ... 
\end{equation} 
such that (i) for any $j \geq 1$, $G_{g_{j+1}}$ is a loop goal of $G_{g_j}$,
and (ii) for no input variable $I$, $D$ contains 
an infinite chain of substitutions for $I$ of the form
\[I/f_1(...,Y_1,...), ..., Y_1/f_2(...,Y_2,...), ..., Y_{i-1}/f_i(...,Y_i,...), ...\]
\end{theorem}

\noindent {\bf Proof:}
($\Longrightarrow$) Assume $MF_{Q_0}$ has an infinite derivation $D'$.
By Theorem \ref{th2}, $GT_{G_0}$ has an infinite derivation $D$
with $D'$ as a moded instance. By Theorem \ref{th1}, $D$ is of form (\ref{eq2})
and satisfies condition (i). 
 
Assume, on the contrary, that $D$ does not satisfy condition (ii).
That is, for some input variable $I$, $D$ contains 
an infinite chain of substitutions for $I$ of the form
\[I/f_1(...,Y_1,...), ..., Y_1/f_2(...,Y_2,...), ..., Y_{i-1}/f_i(...,Y_i,...), ...\] 
Note that for whatever ground term $t$ we assign to $I$, this chain can be instantiated
at most as long in length as the following one: 
\[t/f_1(...,t_1,...), ..., t_1/f_2(...,t_2,...), ..., t_k/f_{k+1}(...,Y_{k+1},...)\] 
where $k = |t|$, $t_i$s are ground terms and $|t_k| = 1$. This means that replacing $I$ with any ground term $t$
leads to a finite moded instance of $D$. Therefore, $D$ has no infinite moded instance in $MF_{Q_0}$,
a contradiction.

($\Longleftarrow$) Assume, on the contrary, that $MF_{Q_0}$ has no infinite 
derivation. By Lemma \ref{lem1}, we reach a contradiction to condition (ii).  
$\Box$

\bigskip

The following corollary is immediate to Theorem \ref{th-main}.

\begin{corollary}[Characterization for a moded query]
\label{cor-main}
A logic program $P$ terminates for a moded query $Q_0$ 
if and only if the moded generalized SLDNF-tree 
$GT_{G_0}$ has no infinite derivation of form (\ref{eq2}) 
satisfying conditions (i) and (ii) of Theorem \ref{th-main}.
\end{corollary}

\begin{example}
\label{eg0-1}
{\em
Consider the moded generalized SLDNF-tree $GT_{G_0}$ in Figure \ref{fig1}.
It has one infinite derivation satisfying condition (i)
of Theorem \ref{th-main}, where for each $j\geq 0$, $N_{g_j} = N_{2j}$.
However, the chain of substitutions for $I$ in this derivation violates condition (ii).
By Corollary \ref{cor-main}, $P_0$ terminates for the moded query $p({\cal I})$.
}
\end{example}

\begin{figure}[htb]
\begin{center}
\setlength{\unitlength}{3947sp}%
\begingroup\makeatletter\ifx\SetFigFont\undefined%
\gdef\SetFigFont#1#2#3#4#5{%
  \reset@font\fontsize{#1}{#2pt}%
  \fontfamily{#3}\fontseries{#4}\fontshape{#5}%
  \selectfont}%
\fi\endgroup%
\begin{picture}(3675,1716)(826,-1006)
\thinlines
{\color[rgb]{0,0,0}\put(2251, 89){\vector( 1, 0){0}}
\multiput(1876, 89)(75.00000,0.00000){5}{\makebox(1.6667,11.6667){\SetFigFont{5}{6}{\rmdefault}{\mddefault}{\updefault}.}}
}%
{\color[rgb]{0,0,0}\put(2776, 14){\vector( 0,-1){300}}
}%
{\color[rgb]{0,0,0}\put(1351,539){\vector( 0,-1){300}}
}%
{\color[rgb]{0,0,0}\put(3901,-436){\vector( 1, 0){0}}
\multiput(3526,-436)(75.00000,0.00000){5}{\makebox(1.6667,11.6667){\SetFigFont{5}{6}{\rmdefault}{\mddefault}{\updefault}.}}
}%
{\color[rgb]{0,0,0}\put(4426,-511){\vector( 0,-1){300}}
}%
\put(4351,-961){\makebox(0,0)[lb]{\smash{\SetFigFont{10}{12.0}{\rmdefault}{\mddefault}{\updefault}{\color[rgb]{0,0,0}$\infty$}%
}}}
\put(1426,389){\makebox(0,0)[lb]{\smash{\SetFigFont{8}{9.6}{\rmdefault}{\mddefault}{\updefault}{\color[rgb]{0,0,0}$\theta_1 = \{X_1/I\}$}%
}}}
\put(2851,-136){\makebox(0,0)[lb]{\smash{\SetFigFont{8}{9.6}{\rmdefault}{\mddefault}{\updefault}{\color[rgb]{0,0,0}$\theta_2 = \{X_2/f(I)\}$}%
}}}
\put(4501,-661){\makebox(0,0)[lb]{\smash{\SetFigFont{8}{9.6}{\rmdefault}{\mddefault}{\updefault}{\color[rgb]{0,0,0}$\theta_3 = \{X_3/f(f(I))\}$}%
}}}
\put(4126,-661){\makebox(0,0)[lb]{\smash{\SetFigFont{8}{9.6}{\rmdefault}{\mddefault}{\updefault}{\color[rgb]{0,0,0}$C_{p_1}$}%
}}}
\put(3976,-436){\makebox(0,0)[lb]{\smash{\SetFigFont{9}{10.8}{\rmdefault}{\mddefault}{\updefault}{\color[rgb]{0,0,0}$N_4$:  $p(f(f(I)))$}%
}}}
\put(826, 89){\makebox(0,0)[lb]{\smash{\SetFigFont{9}{10.8}{\rmdefault}{\mddefault}{\updefault}{\color[rgb]{0,0,0}$N_1$:  $\neg p(f(I))$ }%
}}}
\put(1051,389){\makebox(0,0)[lb]{\smash{\SetFigFont{8}{9.6}{\rmdefault}{\mddefault}{\updefault}{\color[rgb]{0,0,0}$C_{p_1}$}%
}}}
\put(901,614){\makebox(0,0)[lb]{\smash{\SetFigFont{9}{10.8}{\rmdefault}{\mddefault}{\updefault}{\color[rgb]{0,0,0}$N_0$:  $p(I)$}%
}}}
\put(2476,-136){\makebox(0,0)[lb]{\smash{\SetFigFont{8}{9.6}{\rmdefault}{\mddefault}{\updefault}{\color[rgb]{0,0,0}$C_{p_1}$}%
}}}
\put(2326, 89){\makebox(0,0)[lb]{\smash{\SetFigFont{9}{10.8}{\rmdefault}{\mddefault}{\updefault}{\color[rgb]{0,0,0}$N_2$:  $p(f(I))$}%
}}}
\put(2251,-436){\makebox(0,0)[lb]{\smash{\SetFigFont{9}{10.8}{\rmdefault}{\mddefault}{\updefault}{\color[rgb]{0,0,0}$N_3$:  $\neg p(f(f(I)))$ }%
}}}
\end{picture}
\end{center} 
\caption{The moded generalized SLDNF-tree $GT_{\leftarrow p(I)}$.}\label{fig-nonstop} 
\end{figure}
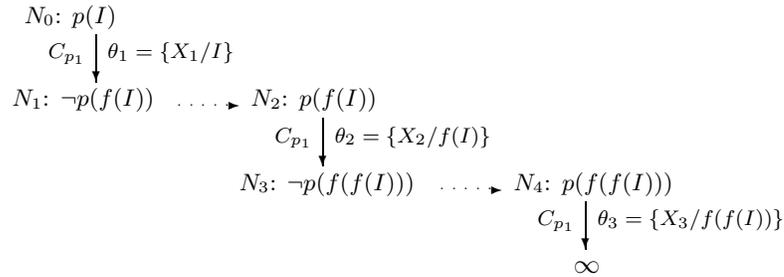

\begin{example} 
\label{eg-nonstop} 
{\em  
Consider the following logic program:
\begin{tabbing}
$\qquad$ \= $P_1:$ $\quad$ \= $q(a).$ \`$C_{q_1}$\\
\> \>            $p(X)\leftarrow \neg p(f(X))$. \`$C_{p_1}$
\end{tabbing}
For a moded query $p({\cal I})$, the moded generalized SLDNF-tree $GT_{\leftarrow p(I)}$ 
is shown in Figure \ref{fig-nonstop}, where $\infty$ represents an infinite extension.
Note that the input variable $I$ is allowed to appear in negative subgoals. 
The infinite derivation in $GT_{\leftarrow p(I)}$ 
satisfies both condition (i) and condition (ii) of Theorem \ref{th-main},
where for each $j\geq 0$, $N_{g_j} = N_{2j}$. 
By Corollary \ref{cor-main}, $P_1$ does not terminate for $p({\cal I})$.
}  
\end{example} 

\section{Testing Termination of Logic Programs for Moded Queries}
\label{sec4}
\subsection{A General Algorithm}
We develop an algorithm for checking termination of logic programs for
moded queries based on Corollary \ref{cor-main}. We begin by
introducing a loop checking mechanism. 

A loop checking mechanism, or more formally a {\em loop check} \cite{BAK91},
defines conditions for us to cut a (possibly infinite) derivation at some node. 
Informally, a loop check is said to be {\em weakly sound} if for any
generalized SLDNF-tree $GT_{G_0}$, $GT_{G_0}$ having a success derivation before cut
implies it has a success derivation after cut; it is said to be {\em complete}
if it cuts all infinite derivations in $GT_{G_0}$. Note that there exists no loop check
that is both weakly sound and complete \cite{BAK91}. In this paper,
we focus on complete loop checks because we want to apply them to
test termination of logic programs.

\begin{definition}
\label{mq-check} 
{\em 
Given a repetition number $r\geq 3$, {\em LP-check} is defined as follows:
any derivation $D$ in $GT_{G_0}$ is cut at 
a node $N_{g_r}$ if $D$ has a partial derivation
\begin{equation} 
\label{eq3}
N_0:G_0\Rightarrow_{C_0} ... N_{g_1}:G_{g_1}\Rightarrow_{C_k}   
...N_{g_2}:G_{g_2}\Rightarrow_{C_k} ... N_{g_r}:G_{g_r}\Rightarrow_{C_k} ... 
\end{equation} 
such that 
(a) for any $j < r$, $G_{g_{j+1}}$ is a loop goal of $G_{g_j}$, 
and (b) for all $j \leq r$, the clause $C_k$ applied to $G_{g_j}$ is the same.  
} 
\end{definition} 

\noindent {\bf Remark:} (1) The repetition number $r$ specifies the minimum
number of loop goals required for a derivation to be cut.
(2) By cutting a derivation at a node $N$ we mean removing all descendants of $N$.

\begin{theorem} 
\label{check-comp} 
LP-check is a complete loop check.
\end{theorem} 
 
\noindent {\bf Proof:} Let $D$ be an infinite derivation in $GT_{G_0}$.
By Theorem \ref{th1}, $D$ is of the form
\[N_0:G_0\Rightarrow_{C_0} ... N_{f_1}:G_{f_1}\Rightarrow_{C_1}   
...N_{f_2}:G_{f_2}\Rightarrow_{C_2} ... \] 
such that for any $i \geq 1$, $G_{f_{i+1}}$ is a loop goal of $G_{f_i}$.
Since a logic program has only a finite number of clauses,
there must be a clause $C_k$ being repeatedly applied at infinitely many nodes 
$N_{g_1}:G_{g_1}, N_{g_2}:G_{g_2}, \cdots$ 
where for each $j \geq 1$, $g_j\in \{f_1, f_2, ...\}$.
Then for any $r>0$, $D$ has a partial derivation of form (\ref{eq3}). So $D$ will be cut
at node $N_{g_r}:G_{g_r}$. This shows
that any infinite derivation can be cut by LP-check. 
That is, LP-check is a complete loop check. $\Box$

\begin{example}[Example \ref{eg1} continued]
\label{eg1-1}
{\em
Let us choose $r=3$ and consider the infinite derivation $D$ in 
Figure \ref{fig1}. $p(X_4)$ at $N_4$ is a loop goal of $p(X_2)$ at $N_2$ 
that is a loop goal of $p(I)$ at $N_0$. Moreover, the same clause
$C_{p_2}$ is applied at the three nodes. $D$  
satisfies the conditions of LP-check and is cut at node $N_4$.
}
\end{example}

We want to apply LP-check to determine termination of logic programs
for moded queries. Recall that to prove that a logic program $P$ terminates
for a moded query $Q_0 = p({\cal I}_1, ..., {\cal I}_m,T_1, ..., T_n)$ is to prove that 
$P$ terminates for any query $p(t_1, ..., t_m,$ $T_2, ..., T_n)$
where each $t_i$ is a ground term. This can be reformulated in
terms of a moded-query forest, that is, $P$ terminates for $Q_0$ if 
$MF_{Q_0}$ has no infinite derivations. Then, Corollary \ref{cor-main}
shows that $P$ terminates for $Q_0$ if the moded generalized SLDNF-tree $GT_{G_0}$ 
has no infinite derivation $D$ of form (\ref{eq2}) that satisfies
the two conditions (i) and (ii). Although this characterization cannot be 
directly used for automated termination test because it requires generating
infinite derivations in $GT_{G_0}$, it can be used together with  
LP-check, as LP-check is able to guess
if a partial derivation would extend to an infinite one.
Before describing our termination testing algorithm, we prepare one more condition 
for Definition \ref{mq-check} based on condition (ii) of Theorem~\ref{th-main}.

\bigskip

\noindent {\bf Condition (c'):}
For no input variable $I$ in $G_{g_1}$, $I$ is recursively substituted
by at least one function via a chain of substitutions 
from $N_{g_1}$ down to $N_{g_r}$.

\bigskip

For instance, in Figure \ref{fig1}, $I$ is recursively substituted
by $f(X_2)$ and $f(X_4)$ via a chain of substitutions 
$I/f(X_2)$, $X_2/f(X_4)$ from $N_0$ down to $N_4$.

Observe that LP-check and Condition (c') implement conditions (i) and (ii)
of Theorem \ref{th-main}, respectively. Although the implementation is not complete
in that it guesses an infinite extension (\ref{eq2}) 
from a partial derivation (\ref{eq3}), such a guess is most likely correct
because it makes full use of the key features (conditions (i) and (ii)
of Theorem \ref{th-main}) of an infinite derivation. This motivates 
the following algorithm.   

\begin{algorithm} 
\label{alg1} 
{\em 
Testing termination of a logic program $P$ for a moded query $Q_0$,
given a repetition number $r\geq 3$. 
\begin{enumerate} 
\item
Initially, set $L = 0$. 
Construct the moded generalized SLDNF-tree $GT_{G_0}$, where for each 
partial derivation $D$ satisfying the conditions of LP-check,
if $D$ satisfies Condition (c') then goto \ref{no}, else set $L = 1$
and extend $D$ with clause $C_k$ skipped.

\item
Return {\em terminating} if $L = 0$; otherwise return {\em most likely terminating}  

\item
\label{no}
Return {\em most likely non-terminating}. 
\end{enumerate} 
} 
\end{algorithm} 

Starting from the root node $N_0:G_0$, we generate 
derivations of a moded generalized SLDNF-tree $GT_{G_0}$ 
step by step. If a partial derivation $D$ of the form
\[N_0:G_0\Rightarrow_{C_0} ... N_{g_1}:G_{g_1}\Rightarrow_{C_k}   
...N_{g_2}:G_{g_2}\Rightarrow_{C_k} ... N_{g_r}:G_{g_r}\]
is generated, which satisfies the conditions of LP-check, then $D$ is 
most likely to extend infinitely in $GT_{G_0}$ (via clause $C_k$). 
By Theorem \ref{th2}, however, $D$ may not have infinite moded instances
in $MF_{Q_0}$. So in this case, we further check $D$ against Condition (c'). 
If Condition (c') is satisfied, we think that $D$ is most likely to 
have moded instances that extend infinitely 
in $MF_{Q_0}$. Algorithm \ref{alg1} then returns 
{\em most likely non-terminating} for $Q_0$.
Otherwise, we continue to extend $D$ by
applying a new clause $C_l$ ($l\neq k$) to $G_{g_r}$ 
($C_k$ is skipped to avoid possible infinite extension).
After all derivations are generated, we distinguish between two cases:
if no derivation was cut by LP-check (i.e. there was no partial derivation $D$ 
satisfying the conditions of LP-check), Algorithm \ref{alg1}  
returns {\em terminating} for $Q_0$;
otherwise, some derivations were cut by LP-check ($L = 1$), so Algorithm \ref{alg1}  
returns {\em most likely terminating} for $Q_0$.

\bigskip 

\noindent {\bf Remark:}
Since a concrete query could be viewed as a special moded query
containing no input variables, Algorithm \ref{alg1} applies to concrete queries as well.
For a concrete query $Q_0$, Condition (c') holds for any derivations.
Therefore, Algorithm \ref{alg1} returns {\em most likely non-terminating} for $Q_0$ 
once a derivation satisfying the conditions of LP-check is generated.

\begin{theorem}
\label{th-terminating}
$P$ terminates for $Q_0$ if Algorithm \ref{alg1} returns an answer {\em terminating}.
\end{theorem}

\noindent {\bf Proof:} If Algorithm \ref{alg1} returns {\em terminating}, 
no derivations were cut by LP-check, so
the moded generalized SLDNF-tree $GT_{G_0}$ for $Q_0$ is finite.
By Theorem \ref{th2}, $MF_{Q_0}$ has no infinite derivation
and thus $P$ terminates for $Q_0$. $\Box$

\bigskip

Algorithm \ref{alg1} applies LP-check to cut possible infinite derivations
in $GT_{G_0}$. Since LP-check is a complete loop check, it cuts all
infinite derivations at some depth. This means that $GT_{G_0}$ after
cut by LP-check is finite. Therefore, Algorithm \ref{alg1} always terminates.

Let $pred(P)$ be the set of predicate symbols in $P$.
Define
\begin{tabbing}
$\quad$ $MQ(P) = \{p({\cal I}_1, ..., {\cal I}_m, X_{m+1},...,X_n) | p$ is a $n$-ary predicate symbol in $pred(P)\}$.
\end{tabbing}
Note that $MQ(P)$ contains all most general moded queries
of $P$ in the sense that any moded query of $P$ is an instance of
some query in $MQ(P)$. 
Since $pred(P)$ is finite, $MQ(P)$ is finite.
Therefore, we can test termination of $P$ for all moded queries 
by applying Algorithm \ref{alg1}  to (a subset of) $MQ(P)$.

\begin{theorem} 
\label{th-inc}
For any two moded queries $Q_0^1 = p({\cal I}_1, ..., {\cal I}_l, X_{l+1},...,X_n)$ 
and $Q_0^2 = p({\cal I}_1, ..., {\cal I}_m, X_{m+1},...,X_n)$ with $l<m$, 
if Algorithm \ref{alg1} returns an answer {\em terminating} 
(resp. {\em most likely terminating}) for $Q_0^1$,  
it returns an answer {\em terminating} (resp. {\em most likely terminating}) for $Q_0^2$. 
\end{theorem} 
 
\noindent {\bf Proof:} 
Any derivation in $GT_{G_0^2}$ that satisfies the conditions of LP-check and
Condition (c') must appear in $GT_{G_0^1}$ and satisfy the conditions.
If Algorithm \ref{alg1} returns {\em terminating} for $Q_0^1$, $GT_{G_0^1}$ is the same
as $GT_{G_0^2}$ with no derivations cut by LP-check. In this case,  
it returns {\em terminating} for $Q_0^2$. 
If Algorithm \ref{alg1} returns {\em most likely terminating} for $Q_0^1$,
$GT_{G_0^1}$ has derivations cut by LP-check, but none of which
satisfies Condition (c'). In this case, $GT_{G_0^2}$ has derivations 
cut by LP-check, none of which satisfies Condition (c'). Therefore, 
Algorithm \ref{alg1} returns {\em most likely terminating} for $Q_0^2$. $\Box$

\bigskip

We use five representative examples to illustrate the 
effectiveness of Algorithm \ref{alg1} (interested readers are
encouraged to apply the algorithm to other benchmark programs).
For each logic program $P_i$, we expect that if $P_i$ terminates for
a query $Q_0$, then Algorithm \ref{alg1} returns {\em terminating} 
or {\em most likely terminating} for $Q_0$, else it returns 
{\em most likely non-terminating} for $Q_0$. 
Let us choose a repetition number $r = 3$.

\begin{example}[Example \ref{eg1-1} continued]
\label{eg1-2}
{\em
Since the partial derivation (Figure \ref{fig1}) between $N_0$ and $N_4$
satisfies the conditions of LP-check, Algorithm \ref{alg1} 
expects that the derivation is most likely to
extend infinitely in $GT_{G_0}$. It then
checks against Condition (c') to see 
if it has moded instances that would extend infinitely in $MF_{Q_0}$.
Clearly, Condition (c') is not satisfied. So Algorithm \ref{alg1} 
skips $C_{p_2}$ and tries to
get a new clause (not yet applied at $N_4$) to expand $N_4$.
Since no new clause is available for $N_4$ and all derivations
of $GT_{G_0}$ except those being cut by LP-check have been
generated, Algorithm \ref{alg1} returns {\em most likely terminating}
for $Q_0$.     
}
\end{example}

\begin{example}[Example \ref{eg-nonstop} continued]
\label{eg-nonstop-2} 
{\em  
The partial derivation between $N_0$ and $N_4$
satisfies both the conditions of LP-check
and Condition (c'), so Algorithm \ref{alg1} returns {\em most likely non-terminating}
for $p({\cal I})$.  
}  
\end{example}

\begin{figure}[htb]
\begin{center}
\setlength{\unitlength}{3947sp}%
\begingroup\makeatletter\ifx\SetFigFont\undefined%
\gdef\SetFigFont#1#2#3#4#5{%
  \reset@font\fontsize{#1}{#2pt}%
  \fontfamily{#3}\fontseries{#4}\fontshape{#5}%
  \selectfont}%
\fi\endgroup%
\begin{picture}(1512,1557)(2776,-847)
\thinlines
{\color[rgb]{0,0,0}\put(3601,-211){\vector(-2,-3){0}}
\multiput(3751, 14)(-37.50000,-56.25000){4}{\makebox(1.6667,11.6667){\SetFigFont{5}{6}{\rmdefault}{\mddefault}{\updefault}.}}
}%
{\color[rgb]{0,0,0}\put(3301,-436){\vector(-2,-3){150}}
}%
{\color[rgb]{0,0,0}\put(3601,-436){\vector( 2,-3){150}}
}%
{\color[rgb]{0,0,0}\put(4126, 14){\vector( 2,-3){150}}
}%
\put(3601,614){\makebox(0,0)[lb]{\smash{\SetFigFont{9}{10.8}{\rmdefault}{\mddefault}{\updefault}{\color[rgb]{0,0,0}$N_0$:  $p(\underline{I})$}%
}}}
\put(4051,389){\makebox(0,0)[lb]{\smash{\SetFigFont{8}{9.6}{\rmdefault}{\mddefault}{\updefault}{\color[rgb]{0,0,0}$\theta_0 = \{X/\underline{I}\}$}%
}}}
\put(3676,389){\makebox(0,0)[lb]{\smash{\SetFigFont{8}{9.6}{\rmdefault}{\mddefault}{\updefault}{\color[rgb]{0,0,0}$C_{p_1}$}%
}}}
\put(2776,-811){\makebox(0,0)[lb]{\smash{\SetFigFont{9}{10.8}{\rmdefault}{\mddefault}{\updefault}{\color[rgb]{0,0,0}$N_3$:  $\Box_t$}%
}}}
\put(3526,-811){\makebox(0,0)[lb]{\smash{\SetFigFont{9}{10.8}{\rmdefault}{\mddefault}{\updefault}{\color[rgb]{0,0,0}$N_4$:  $q$}%
}}}
\put(3226,-361){\makebox(0,0)[lb]{\smash{\SetFigFont{9}{10.8}{\rmdefault}{\mddefault}{\updefault}{\color[rgb]{0,0,0}$N_2$:  $q$}%
}}}
\put(3676,-511){\makebox(0,0)[lb]{\smash{\SetFigFont{8}{9.6}{\rmdefault}{\mddefault}{\updefault}{\color[rgb]{0,0,0}$C_{q_2}$}%
}}}
\put(2926,-511){\makebox(0,0)[lb]{\smash{\SetFigFont{8}{9.6}{\rmdefault}{\mddefault}{\updefault}{\color[rgb]{0,0,0}$C_{q_1}$}%
}}}
\put(4126,-361){\makebox(0,0)[lb]{\smash{\SetFigFont{9}{10.8}{\rmdefault}{\mddefault}{\updefault}{\color[rgb]{0,0,0}$N_5:\Box_f$}%
}}}
\put(3601, 89){\makebox(0,0)[lb]{\smash{\SetFigFont{9}{10.8}{\rmdefault}{\mddefault}{\updefault}{\color[rgb]{0,0,0}$N_1$:  $\neg q$ }%
}}}
{\color[rgb]{0,0,0}\put(3976,539){\vector( 0,-1){300}}
}%
\end{picture}
\end{center} 
\caption{A moded generalized SLDNF-tree $GT_{\leftarrow p(I)}$}\label{fig-p2} 
\end{figure}
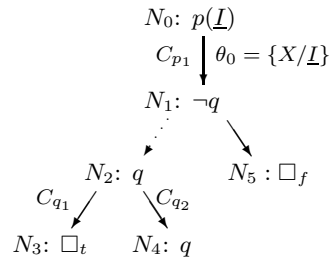

\begin{example} 
\label{eg-p2} 
{\em 
Consider the following logic program:
\begin{tabbing} 
$\qquad$ \= $P_2:$ $\quad$ \= $p(X)\leftarrow \neg q$. \`$C_{p_1}$\\ 
\>\> $q$.       \`$C_{q_1}$ \\ 
\>\> $q \leftarrow q$.            \`$C_{q_2}$ 
\end{tabbing} 
For a moded query $p({\cal I})$, Algorithm \ref{alg1} 
generates a moded generalized SLDNF-tree $GT_{\leftarrow p(I)}$, as depicted in 
Figure \ref{fig-p2} where input variables are underlined.\footnote{Note
that the subsidiary SLDNF$^*$-tree rooted at $N_2:q$ for $\neg q$
terminates at the first success leaf $N_3$, so $N_4$ is not extended.}
Since no derivation is cut by LP-check,
Algorithm \ref{alg1} returns {\em terminating} for $p({\cal I})$. 
}  
\end{example}   

\begin{figure}[htb]
\begin{center}
\setlength{\unitlength}{3947sp}%
\begingroup\makeatletter\ifx\SetFigFont\undefined%
\gdef\SetFigFont#1#2#3#4#5{%
  \reset@font\fontsize{#1}{#2pt}%
  \fontfamily{#3}\fontseries{#4}\fontshape{#5}%
  \selectfont}%
\fi\endgroup%
\begin{picture}(4650,1521)(826,-811)
\put(1126,-811){\makebox(0,0)[lb]{\smash{\SetFigFont{10}{12.0}{\rmdefault}{\mddefault}{\updefault}{\color[rgb]{0,0,0}(a)}%
}}}
\put(3226,-811){\makebox(0,0)[lb]{\smash{\SetFigFont{10}{12.0}{\rmdefault}{\mddefault}{\updefault}{\color[rgb]{0,0,0}(b)}%
}}}
\put(5326,-811){\makebox(0,0)[lb]{\smash{\SetFigFont{10}{12.0}{\rmdefault}{\mddefault}{\updefault}{\color[rgb]{0,0,0}(c)}%
}}}
\put(3376,-61){\makebox(0,0)[lb]{\smash{\SetFigFont{7}{8.4}{\rmdefault}{\mddefault}{\updefault}{\color[rgb]{0,0,0}$\theta_1 = \{Y/[X_1|Y_1],$}%
}}}
\put(3001,-136){\makebox(0,0)[lb]{\smash{\SetFigFont{7}{8.4}{\rmdefault}{\mddefault}{\updefault}{\color[rgb]{0,0,0}$C_{a_2}$}%
}}}
\put(3376,464){\makebox(0,0)[lb]{\smash{\SetFigFont{7}{8.4}{\rmdefault}{\mddefault}{\updefault}{\color[rgb]{0,0,0}$\theta_0 = \{V_1/[X|Y],$}%
}}}
\put(3376,-211){\makebox(0,0)[lb]{\smash{\SetFigFont{7}{8.4}{\rmdefault}{\mddefault}{\updefault}{\color[rgb]{0,0,0}$U_1/\underline{I},Z/[X_1|Z_1]\}$}%
}}}
\put(3376,314){\makebox(0,0)[lb]{\smash{\SetFigFont{7}{8.4}{\rmdefault}{\mddefault}{\updefault}{\color[rgb]{0,0,0}$U/\underline{I},V_3/[X|Z]\}$}%
}}}
\thinlines
{\color[rgb]{0,0,0}\put(5401,539){\vector( 0,-1){300}}
}%
{\color[rgb]{0,0,0}\put(5401, 14){\vector( 0,-1){300}}
}%
\put(5026, 89){\makebox(0,0)[lb]{\smash{\SetFigFont{9}{10.8}{\rmdefault}{\mddefault}{\updefault}{\color[rgb]{0,0,0}$N_1$:  $append(Y,V_2,\underline{Z})$ }%
}}}
\put(5026,-436){\makebox(0,0)[lb]{\smash{\SetFigFont{9}{10.8}{\rmdefault}{\mddefault}{\updefault}{\color[rgb]{0,0,0}$N_2$:  $append(Y_1,V_2,\underline{Z_1})$ }%
}}}
\put(5026,614){\makebox(0,0)[lb]{\smash{\SetFigFont{9}{10.8}{\rmdefault}{\mddefault}{\updefault}{\color[rgb]{0,0,0}$N_0$:  $append(V_1, V_2, \underline{I})$}%
}}}
\put(5101,389){\makebox(0,0)[lb]{\smash{\SetFigFont{7}{8.4}{\rmdefault}{\mddefault}{\updefault}{\color[rgb]{0,0,0}$C_{a_2}$}%
}}}
\put(5101,-136){\makebox(0,0)[lb]{\smash{\SetFigFont{7}{8.4}{\rmdefault}{\mddefault}{\updefault}{\color[rgb]{0,0,0}$C_{a_2}$}%
}}}
\put(5476,-61){\makebox(0,0)[lb]{\smash{\SetFigFont{7}{8.4}{\rmdefault}{\mddefault}{\updefault}{\color[rgb]{0,0,0}$\theta_1 = \{Y/[X_1|Y_1],$}%
}}}
\put(5476,464){\makebox(0,0)[lb]{\smash{\SetFigFont{7}{8.4}{\rmdefault}{\mddefault}{\updefault}{\color[rgb]{0,0,0}$\theta_0 = \{V_1/[X|Y],$}%
}}}
\put(5476,-211){\makebox(0,0)[lb]{\smash{\SetFigFont{7}{8.4}{\rmdefault}{\mddefault}{\updefault}{\color[rgb]{0,0,0}$U_1/V_2,\underline{Z}/[X_1|Z_1]\}$}%
}}}
\put(5476,314){\makebox(0,0)[lb]{\smash{\SetFigFont{7}{8.4}{\rmdefault}{\mddefault}{\updefault}{\color[rgb]{0,0,0}$U/V_2,\underline{I}/[X|Z]\}$}%
}}}
\put(3001,389){\makebox(0,0)[lb]{\smash{\SetFigFont{7}{8.4}{\rmdefault}{\mddefault}{\updefault}{\color[rgb]{0,0,0}$C_{a_2}$}%
}}}
{\color[rgb]{0,0,0}\put(1201,539){\vector( 0,-1){300}}
}%
{\color[rgb]{0,0,0}\put(1201, 14){\vector( 0,-1){300}}
}%
\put(826, 89){\makebox(0,0)[lb]{\smash{\SetFigFont{9}{10.8}{\rmdefault}{\mddefault}{\updefault}{\color[rgb]{0,0,0}$N_1$:  $append(\underline{Y},V_2,Z)$ }%
}}}
\put(826,-436){\makebox(0,0)[lb]{\smash{\SetFigFont{9}{10.8}{\rmdefault}{\mddefault}{\updefault}{\color[rgb]{0,0,0}$N_2$:  $append(\underline{Y_1},V_2,Z_1)$ }%
}}}
\put(826,614){\makebox(0,0)[lb]{\smash{\SetFigFont{9}{10.8}{\rmdefault}{\mddefault}{\updefault}{\color[rgb]{0,0,0}$N_0$:  $append(\underline{I}, V_2, V_3)$}%
}}}
\put(901,389){\makebox(0,0)[lb]{\smash{\SetFigFont{7}{8.4}{\rmdefault}{\mddefault}{\updefault}{\color[rgb]{0,0,0}$C_{a_2}$}%
}}}
\put(901,-136){\makebox(0,0)[lb]{\smash{\SetFigFont{7}{8.4}{\rmdefault}{\mddefault}{\updefault}{\color[rgb]{0,0,0}$C_{a_2}$}%
}}}
\put(1276,-61){\makebox(0,0)[lb]{\smash{\SetFigFont{7}{8.4}{\rmdefault}{\mddefault}{\updefault}{\color[rgb]{0,0,0}$\theta_1 = \{\underline{Y}/[X_1|Y_1],$}%
}}}
\put(1276,464){\makebox(0,0)[lb]{\smash{\SetFigFont{7}{8.4}{\rmdefault}{\mddefault}{\updefault}{\color[rgb]{0,0,0}$\theta_0 = \{\underline{I}/[X|Y],$}%
}}}
\put(1276,-211){\makebox(0,0)[lb]{\smash{\SetFigFont{7}{8.4}{\rmdefault}{\mddefault}{\updefault}{\color[rgb]{0,0,0}$U_1/V_2,Z/[X_1|Z_1]\}$}%
}}}
\put(1276,314){\makebox(0,0)[lb]{\smash{\SetFigFont{7}{8.4}{\rmdefault}{\mddefault}{\updefault}{\color[rgb]{0,0,0}$U/V_2,V_3/[X|Z]\}$}%
}}}
{\color[rgb]{0,0,0}\put(3301,539){\vector( 0,-1){300}}
}%
{\color[rgb]{0,0,0}\put(3301, 14){\vector( 0,-1){300}}
}%
\put(2926, 89){\makebox(0,0)[lb]{\smash{\SetFigFont{9}{10.8}{\rmdefault}{\mddefault}{\updefault}{\color[rgb]{0,0,0}$N_1$:  $append(Y,\underline{I},Z)$ }%
}}}
\put(2926,-436){\makebox(0,0)[lb]{\smash{\SetFigFont{9}{10.8}{\rmdefault}{\mddefault}{\updefault}{\color[rgb]{0,0,0}$N_2$:  $append(Y_1,\underline{I},Z_1)$ }%
}}}
\put(2926,614){\makebox(0,0)[lb]{\smash{\SetFigFont{9}{10.8}{\rmdefault}{\mddefault}{\updefault}{\color[rgb]{0,0,0}$N_0$:  $append(V_1, \underline{I}, V_3)$}%
}}}
\end{picture}
\end{center} 
\caption{(a) $GT_{G_0^1}$, (b) $GT_{G_0^2}$, and (c) $GT_{G_0^3}$}\label{fig2} 
\end{figure}
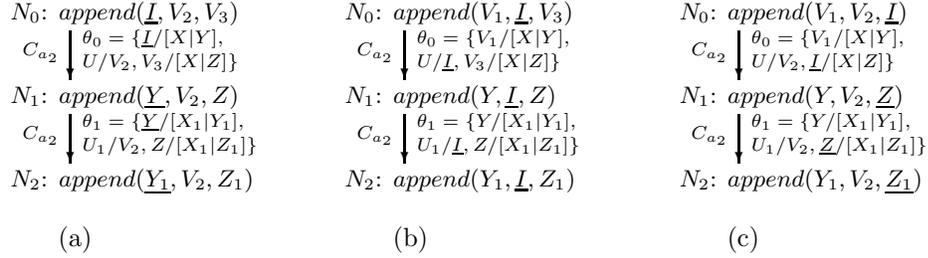

\begin{example} 
\label{eg2} 
{\em 
Consider the following logic program:
\begin{tabbing} 
$\qquad$ \= $P_3:$ $\quad$ \= $append([],X,X)$. \`$C_{a_1}$\\ 
\>\> $append([X|Y],U,[X|Z])\leftarrow append(Y,U,Z)$.       \`$C_{a_2}$  
\end{tabbing} 
Let us choose the three simplest moded queries:  
\begin{tabbing} 
$\qquad\qquad$ \= $Q_0^1= append({\cal I}, V_2, V_3)$,\\ 
\>                $Q_0^2= append(V_1, {\cal I}, V_3)$,\\ 
\>                $Q_0^3= append(V_1, V_2, {\cal I})$.
\end{tabbing} 
Since applying clause $C_{a_1}$ produces only leaf nodes,
for simplicity we ignore it when depicting moded generalized
SLDNF-trees. It is quite easy to 
determine the termination behavior for the above three moded queries. 
Algorithm \ref{alg1} builds $GT_{G_0^1}$, $GT_{G_0^2}$ and $GT_{G_0^3}$
as shown in Figures \ref{fig2} (a), (b) and (c), respectively. Note that all the derivations
starting at $N_0$ and ending at $N_2$ satisfy the conditions of LP-check, so 
they are cut at $N_2$. Since the derivations in $GT_{G_0^1}$ and $GT_{G_0^3}$
do not satisfy Condition (c') ($I$ is recursively substituted via a chain of substitutions
$\underline{I}/[X|Y], \underline{Y}/[X_1|Y_1]$), 
Algorithm \ref{alg1} returns {\em most likely terminating}
for $Q_0^1$ and $Q_0^3$. Since the derivation in $GT_{G_0^2}$ satisfies
Condition (c'), Algorithm \ref{alg1} returns {\em most likely non-terminating} for $Q_0^2$. 
By Theorem \ref{th-inc}, we infer that $P_3$ most likely terminates 
for all moded queries in $MQ(P_3)$ except for $Q_0^2$.
}
\end{example} 
 
\begin{figure}[htb]
\setlength{\unitlength}{3947sp}%
\begingroup\makeatletter\ifx\SetFigFont\undefined%
\gdef\SetFigFont#1#2#3#4#5{%
  \reset@font\fontsize{#1}{#2pt}%
  \fontfamily{#3}\fontseries{#4}\fontshape{#5}%
  \selectfont}%
\fi\endgroup%
\begin{picture}(4350,3555)(1276,-3361)
\put(5626,-3361){\makebox(0,0)[lb]{\smash{\SetFigFont{10}{12.0}{\rmdefault}{\mddefault}{\updefault}{\color[rgb]{0,0,0}(b)}%
}}}
\put(1801,-3361){\makebox(0,0)[lb]{\smash{\SetFigFont{10}{12.0}{\rmdefault}{\mddefault}{\updefault}{\color[rgb]{0,0,0}(a)}%
}}}
\put(1351,-136){\makebox(0,0)[lb]{\smash{\SetFigFont{7}{8.4}{\rmdefault}{\mddefault}{\updefault}{\color[rgb]{0,0,0}$C_{m_1}$}%
}}}
\put(1351,-661){\makebox(0,0)[lb]{\smash{\SetFigFont{7}{8.4}{\rmdefault}{\mddefault}{\updefault}{\color[rgb]{0,0,0}$C_{m_1}$}%
}}}
\put(1726,-136){\makebox(0,0)[lb]{\smash{\SetFigFont{7}{8.4}{\rmdefault}{\mddefault}{\updefault}{\color[rgb]{0,0,0}$\theta_0 = \{\underline{I}/s(X_1),Y_1/V_2,Z_1/V_3\}$}%
}}}
\put(1726,-661){\makebox(0,0)[lb]{\smash{\SetFigFont{7}{8.4}{\rmdefault}{\mddefault}{\updefault}{\color[rgb]{0,0,0}$\theta_1 = \{\underline{X_1}/s(X_2),Y_2/V_2,Z_2/U_1\}$}%
}}}
\put(5101, 89){\makebox(0,0)[lb]{\smash{\SetFigFont{9}{10.8}{\rmdefault}{\mddefault}{\updefault}{\color[rgb]{0,0,0}$N_0$:  $mult(\underline{I_1}, \underline{I_2}, V_3)$}%
}}}
\put(5101,-3061){\makebox(0,0)[lb]{\smash{\SetFigFont{9}{10.8}{\rmdefault}{\mddefault}{\updefault}{\color[rgb]{0,0,0}$N_6$:  $add(\underline{X_4},s(s(\underline{X_4})),Z_4)$ }%
}}}
\put(5176,-2236){\makebox(0,0)[lb]{\smash{\SetFigFont{7}{8.4}{\rmdefault}{\mddefault}{\updefault}{\color[rgb]{0,0,0}$C_{a_1}$}%
}}}
\put(5176,-2761){\makebox(0,0)[lb]{\smash{\SetFigFont{7}{8.4}{\rmdefault}{\mddefault}{\updefault}{\color[rgb]{0,0,0}$C_{a_1}$}%
}}}
\put(5551,-2236){\makebox(0,0)[lb]{\smash{\SetFigFont{7}{8.4}{\rmdefault}{\mddefault}{\updefault}{\color[rgb]{0,0,0}$\theta_4 = \{\underline{I_2}/s(X_3),V_3/s(Z_3)\}$}%
}}}
\put(5551,-2761){\makebox(0,0)[lb]{\smash{\SetFigFont{7}{8.4}{\rmdefault}{\mddefault}{\updefault}{\color[rgb]{0,0,0}$\theta_5 = \{\underline{X_3}/s(X_4),Z_3/s(Z_4)\}$}%
}}}
\put(5101,-1486){\makebox(0,0)[lb]{\smash{\SetFigFont{9}{10.8}{\rmdefault}{\mddefault}{\updefault}{\color[rgb]{0,0,0}$N_3$:  $add(0,\underline{I_2},U_1),add(U_1,\underline{I_2},V_3)$ }%
}}}
\put(5101,-2011){\makebox(0,0)[lb]{\smash{\SetFigFont{9}{10.8}{\rmdefault}{\mddefault}{\updefault}{\color[rgb]{0,0,0}$N_4$:  $add(\underline{I_2},\underline{I_2},V_3)$ }%
}}}
\put(5176,-1186){\makebox(0,0)[lb]{\smash{\SetFigFont{7}{8.4}{\rmdefault}{\mddefault}{\updefault}{\color[rgb]{0,0,0}$C_{m_2}$}%
}}}
\put(5176,-1711){\makebox(0,0)[lb]{\smash{\SetFigFont{7}{8.4}{\rmdefault}{\mddefault}{\updefault}{\color[rgb]{0,0,0}$C_{a_2}$}%
}}}
\put(5551,-1186){\makebox(0,0)[lb]{\smash{\SetFigFont{7}{8.4}{\rmdefault}{\mddefault}{\updefault}{\color[rgb]{0,0,0}$\theta_2 = \{\underline{X_2}/0,Y_3/\underline{I_2},U_2/0\}$}%
}}}
\put(5551,-1711){\makebox(0,0)[lb]{\smash{\SetFigFont{7}{8.4}{\rmdefault}{\mddefault}{\updefault}{\color[rgb]{0,0,0}$\theta_3 = \{Y_4/\underline{I_2},U_1/\underline{I_2}\}$}%
}}}
\put(5101,-436){\makebox(0,0)[lb]{\smash{\SetFigFont{9}{10.8}{\rmdefault}{\mddefault}{\updefault}{\color[rgb]{0,0,0}$N_1$:  $mult(\underline{X_1},\underline{I_2},U_1),add(U_1,\underline{I_2},V_3)$ }%
}}}
\put(5101,-961){\makebox(0,0)[lb]{\smash{\SetFigFont{9}{10.8}{\rmdefault}{\mddefault}{\updefault}{\color[rgb]{0,0,0}$N_2$:  $mult(\underline{X_2},\underline{I_2},U_2),add(U_2,\underline{I_2},U_1),add(U_1,\underline{I_2},V_3)$ }%
}}}
\put(5176,-136){\makebox(0,0)[lb]{\smash{\SetFigFont{7}{8.4}{\rmdefault}{\mddefault}{\updefault}{\color[rgb]{0,0,0}$C_{m_1}$}%
}}}
\put(5176,-661){\makebox(0,0)[lb]{\smash{\SetFigFont{7}{8.4}{\rmdefault}{\mddefault}{\updefault}{\color[rgb]{0,0,0}$C_{m_1}$}%
}}}
\put(5551,-136){\makebox(0,0)[lb]{\smash{\SetFigFont{7}{8.4}{\rmdefault}{\mddefault}{\updefault}{\color[rgb]{0,0,0}$\theta_0 = \{\underline{I_1}/s(X_1),Y_1/\underline{I_2},Z_1/V_3\}$}%
}}}
\put(5551,-661){\makebox(0,0)[lb]{\smash{\SetFigFont{7}{8.4}{\rmdefault}{\mddefault}{\updefault}{\color[rgb]{0,0,0}$\theta_1 = \{\underline{X_1}/s(X_2),Y_2/\underline{I_2},Z_2/U_1\}$}%
}}}
\put(5101,-2536){\makebox(0,0)[lb]{\smash{\SetFigFont{9}{10.8}{\rmdefault}{\mddefault}{\updefault}{\color[rgb]{0,0,0}$N_5$:  $add(\underline{X_3},s(\underline{X_3}),Z_3)$ }%
}}}
\put(1276,-961){\makebox(0,0)[lb]{\smash{\SetFigFont{9}{10.8}{\rmdefault}{\mddefault}{\updefault}{\color[rgb]{0,0,0}$N_2$:  $mult(\underline{X_2},V_2,U_2),add(U_2,V_2,U_1),add(U_1,V_2,V_3)$ }%
}}}
\thinlines
{\color[rgb]{0,0,0}\put(1651,-2086){\vector( 0,-1){300}}
}%
{\color[rgb]{0,0,0}\put(1651,-2611){\vector( 0,-1){300}}
}%
{\color[rgb]{0,0,0}\put(1651,-1036){\vector( 0,-1){300}}
}%
{\color[rgb]{0,0,0}\put(1651,-1561){\vector( 0,-1){300}}
}%
{\color[rgb]{0,0,0}\put(1651, 14){\vector( 0,-1){300}}
}%
{\color[rgb]{0,0,0}\put(1651,-511){\vector( 0,-1){300}}
}%
{\color[rgb]{0,0,0}\put(5476,-2086){\vector( 0,-1){300}}
}%
{\color[rgb]{0,0,0}\put(5476,-2611){\vector( 0,-1){300}}
}%
{\color[rgb]{0,0,0}\put(5476,-1036){\vector( 0,-1){300}}
}%
{\color[rgb]{0,0,0}\put(5476,-1561){\vector( 0,-1){300}}
}%
{\color[rgb]{0,0,0}\put(5476, 14){\vector( 0,-1){300}}
}%
{\color[rgb]{0,0,0}\put(5476,-511){\vector( 0,-1){300}}
}%
\put(1276, 89){\makebox(0,0)[lb]{\smash{\SetFigFont{9}{10.8}{\rmdefault}{\mddefault}{\updefault}{\color[rgb]{0,0,0}$N_0$:  $mult(\underline{I}, V_2, V_3)$}%
}}}
\put(1276,-2536){\makebox(0,0)[lb]{\smash{\SetFigFont{9}{10.8}{\rmdefault}{\mddefault}{\updefault}{\color[rgb]{0,0,0}$N_5$:  $add(X_3,s(X_3),Z_3)$ }%
}}}
\put(1276,-3061){\makebox(0,0)[lb]{\smash{\SetFigFont{9}{10.8}{\rmdefault}{\mddefault}{\updefault}{\color[rgb]{0,0,0}$N_6$:  $add(X_4,s(s(X_4)),Z_4)$ }%
}}}
\put(1351,-2236){\makebox(0,0)[lb]{\smash{\SetFigFont{7}{8.4}{\rmdefault}{\mddefault}{\updefault}{\color[rgb]{0,0,0}$C_{a_1}$}%
}}}
\put(1351,-2761){\makebox(0,0)[lb]{\smash{\SetFigFont{7}{8.4}{\rmdefault}{\mddefault}{\updefault}{\color[rgb]{0,0,0}$C_{a_1}$}%
}}}
\put(1726,-2236){\makebox(0,0)[lb]{\smash{\SetFigFont{7}{8.4}{\rmdefault}{\mddefault}{\updefault}{\color[rgb]{0,0,0}$\theta_4 = \{V_2/s(X_3),V_3/s(Z_3)\}$}%
}}}
\put(1726,-2761){\makebox(0,0)[lb]{\smash{\SetFigFont{7}{8.4}{\rmdefault}{\mddefault}{\updefault}{\color[rgb]{0,0,0}$\theta_5 = \{X_3/s(X_4),Z_3/s(Z_4)\}$}%
}}}
\put(1276,-1486){\makebox(0,0)[lb]{\smash{\SetFigFont{9}{10.8}{\rmdefault}{\mddefault}{\updefault}{\color[rgb]{0,0,0}$N_3$:  $add(0,V_2,U_1),add(U_1,V_2,V_3)$ }%
}}}
\put(1276,-2011){\makebox(0,0)[lb]{\smash{\SetFigFont{9}{10.8}{\rmdefault}{\mddefault}{\updefault}{\color[rgb]{0,0,0}$N_4$:  $add(V_2,V_2,V_3)$ }%
}}}
\put(1351,-1186){\makebox(0,0)[lb]{\smash{\SetFigFont{7}{8.4}{\rmdefault}{\mddefault}{\updefault}{\color[rgb]{0,0,0}$C_{m_2}$}%
}}}
\put(1351,-1711){\makebox(0,0)[lb]{\smash{\SetFigFont{7}{8.4}{\rmdefault}{\mddefault}{\updefault}{\color[rgb]{0,0,0}$C_{a_2}$}%
}}}
\put(1726,-1186){\makebox(0,0)[lb]{\smash{\SetFigFont{7}{8.4}{\rmdefault}{\mddefault}{\updefault}{\color[rgb]{0,0,0}$\theta_2 = \{\underline{X_2}/0,Y_3/V_2,U_2/0\}$}%
}}}
\put(1726,-1711){\makebox(0,0)[lb]{\smash{\SetFigFont{7}{8.4}{\rmdefault}{\mddefault}{\updefault}{\color[rgb]{0,0,0}$\theta_3 = \{Y_4/V_2,U_1/V_2\}$}%
}}}
\put(1276,-436){\makebox(0,0)[lb]{\smash{\SetFigFont{9}{10.8}{\rmdefault}{\mddefault}{\updefault}{\color[rgb]{0,0,0}$N_1$:  $mult(\underline{X_1},V_2,U_1),add(U_1,V_2,V_3)$ }%
}}}
\end{picture}
\caption{Two moded generalized SLDNF-trees}\label{fig-eg3} 
\end{figure}
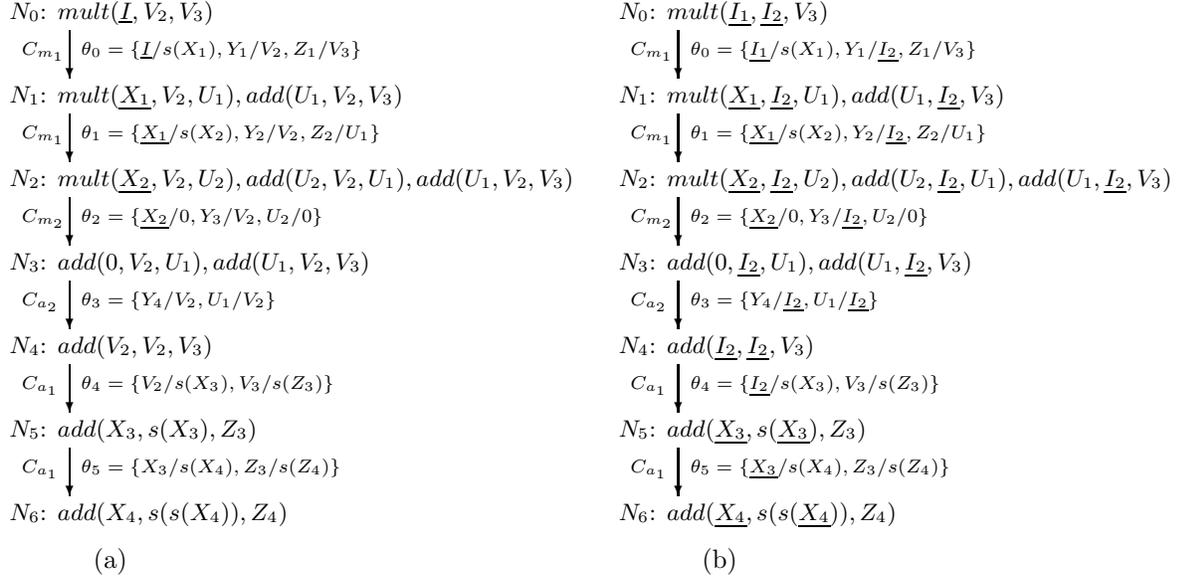

\begin{example} 
\label{eg3} 
{\em 
Consider the following logic program:
\begin{tabbing} 
$\qquad$ \= $P_4:$ $\quad$ \= $mult(s(X),Y,Z)\leftarrow mult(X,Y,U),add(U,Y,Z)$. \`$C_{m_1}$\\ 
\>\>                          $mult(0,Y,0)$.       \`$C_{m_2}$ \\[.06in]
\>\>                          $add(s(X),Y,s(Z))\leftarrow add(X,Y,Z)$.   \`$C_{a_1}$\\ 
\>\>                          $add(0,Y,Y)$.   \`$C_{a_2}$
\end{tabbing}
$MQ(P_4)$ consists of fourteen moded queries, seven for $mult(.)$ and seven for $add(.)$. 
Applying Algorithm \ref{alg1} yields the solution: (1)
$P_4$ most likely terminates for all moded queries of 
$add(.)$ except for $add(V_1,{\cal I}_2,V_3)$
that is most likely non-terminating,
and (2) $P_4$ most likely terminates for $mult({\cal I}_1,{\cal I}_2,V_3)$
and $mult({\cal I}_1,{\cal I}_2,{\cal I}_3)$ but is most likely non-terminating
for the remaining moded queries of $mult(.)$. 
For illustration, we depict two moded generalized
SLDNF-trees for $mult({\cal I},V_2,V_3)$ and $mult({\cal I}_1,{\cal I}_2,V_3)$,
as shown in Figures \ref{fig-eg3} (a) and (b), respectively.
In the two moded generalized SLDNF-trees, the partial derivation from $N_0$ down to $N_2$ satisfies 
the conditions of LP-check but violates Condition (c'), so clause $C_{m_1}$
is skipped when expanding $N_2$. When the derivation is extended to $N_6$,
the conditions of LP-check are satisfied again, where $G_6$ is a loop goal of $G_5$ 
that is a loop goal of $G_4$. Since the derivation for $mult({\cal I},V_2,V_3)$ 
(Figure \ref{fig-eg3} (a)) also satisfies Condition (c'), Algorithm \ref{alg1}
returns an answer $-${\em most likely non-terminating} $-$ for this moded query. 
The derivation for $mult({\cal I}_1,{\cal I}_2,V_3)$ 
(Figure \ref{fig-eg3} (b)) does not satisfy Condition (c'), so clause 
$C_{a_1}$ is skipped to expand $N_6$. For simplicity, we omitted all derivations
leading to a leaf node $\Box_t$. Because there is no derivation satisfying both
the conditions of LP-check and Condition (c'), Algorithm \ref{alg1} ends up with
an answer $-$ {\em most likely terminating} $-$ for $mult({\cal I}_1,{\cal I}_2,V_3)$.
It is then immediately inferred by Theorem \ref{th-inc} 
that $P_4$ most likely terminates for $mult({\cal I}_1,{\cal I}_2,{\cal I}_3)$.
}
\end{example} 

For each of the above five 
example logic programs, $P_0-P_4$, it terminates 
for a moded query if and only if applying 
Algorithm \ref{alg1} with the smallest repetition number 
$r = 3$ yields an answer $-$ {\em terminating} 
or {\em most likely terminating} $-$ for the query. 
This is true for commonly used benchmark logic programs in the literature.
Due to the undecidability of the termination problem, however,
there exist cases that Algorithm \ref{alg1} yields an incorrect answer
unless a big repetition number is used. 
Consider the following carefully created logic program:
\begin{tabbing} 
$\qquad$ \= $P_5:$ $\quad$ \= $p(f(X),Y) \leftarrow  p(X,s(Y)).$ \`$C_{p_1}$\\
\> \>      $p(Z, \underbrace{s(s(...s}_{100 \ items}(0)...))) \leftarrow q.$ \`$C_{p_2}$\\
\>   \>    $q \leftarrow  q.$  \`$C_{q_1}$
\end{tabbing} 
$P_5$ does not terminate for a moded query $Q_0 = p({\cal I}, 0)$, but
Algorithm \ref{alg1} will return {\em most likely terminating} for $Q_0$ 
unless the repetition number $r$ is set above 100. 

The question of which repetition number (also called {\em depth bound} in some literature) 
is optimal remains open for a long
time in loop checking \cite{Bol93,shen001}. In \cite{shen001},
the authors say ``The only way to deal with this problem is by
heuristically tuning the depth bound in practical situations."
However, up till now we see no heuristic methods reported in the literature.

In this paper, we propose a simple yet effective heuristic method for
handling the repetition number problem. Observe that 
due to the large argument $\underbrace{s(s(...s}_{100 \ items}(0)...))$
in its head, the second clause of $P_5$
cannot be applicable to $p(I, 0)$.
However, the second argument of $p(I, 0)$ can grow as large as 
$\underbrace{s(s(...s}_{100 \ items}(0)...))$
if the first clause is repeatedly applied.
Our intuition then is that instead of choosing a big repetition number, 
we use a small one (say $r = 3$) with some additional constraints
that help $p(I, 0)$ grow up to its expected size before a derivation is cut.
For each $n$-ary predicate symbol $p$,
let $p_{max}^i$ ($1\leq i \leq n$) denote the maximum 
layers of nested functions in the $i$-th 
argument of all clause heads $p(.)$. For instance, in $P_5$ 
$p_{max}^1 = 1$ and $p_{max}^2 = 100$. The following heuristic
defines a constraint. 

\bigskip

\noindent {\bf Heuristic 1:} When some arguments of $p(.)$ grow in a sequence of
loop goals, if a derivation is cut at some loop goal, each $i$-th growing argument of $p(.)$  
in this goal has at least $p_{max}^i$ layers of nested functions.

\bigskip

It is easy to enhance LP-check with Heuristic 1, simply by adding a third condition
to Definition \ref{mq-check}:
\begin{enumerate}
\item[(c)] 
Let $L_{g_j}^1 = p(.)$. If some arguments of $p(.)$ grow from $N_{g_1}$, 
$N_{g_2}$, ..., to $N_{g_r}$, then each $i$-th growing argument of $p(.)$ at $N_{g_r}$
has at least $p_{max}^i$ layers of nested functions. 
\end{enumerate} 

It is easy to prove that enhancing LP-check with Heuristic 1 does not change 
the completeness of LP-check. Let $D$ be an infinite derivation 
and let $S$ be the set of finite partial derivations of $D$ satisfying
conditions (a) and (b) of LP-check and satisfying the if-part
of condition (c). Assume, on the contrary, that no derivation in $S$
satisfies the then-part of condition (c) (in this case, $D$ will not
cut by LP-check enhanced with Heuristic 1).  This case will never occur unless
for some $i$-th argument of $p(.)$, $p_{max}^i$ is an infinite number.
Since $p_{max}^i$ is finite, the above assumption does not hold.

\begin{figure}[htb]
\begin{center}
\setlength{\unitlength}{3947sp}%
\begingroup\makeatletter\ifx\SetFigFont\undefined%
\gdef\SetFigFont#1#2#3#4#5{%
  \reset@font\fontsize{#1}{#2pt}%
  \fontfamily{#3}\fontseries{#4}\fontshape{#5}%
  \selectfont}%
\fi\endgroup%
\begin{picture}(450,3582)(1276,-3397)
\thicklines
{\color[rgb]{0,0,0}\multiput(1651,-1036)(0.00000,-60.00000){6}{\makebox(6.6667,10.0000){\SetFigFont{10}{12}{\rmdefault}{\mddefault}{\updefault}.}}
}%
\put(1726,-2536){\makebox(0,0)[lb]{\smash{\SetFigFont{7}{8.4}{\rmdefault}{\mddefault}{\updefault}{\color[rgb]{0,0,0}$C_{q_1}$}%
}}}
\put(1726,-3061){\makebox(0,0)[lb]{\smash{\SetFigFont{7}{8.4}{\rmdefault}{\mddefault}{\updefault}{\color[rgb]{0,0,0}$C_{q_1}$}%
}}}
\put(1726,-2011){\makebox(0,0)[lb]{\smash{\SetFigFont{7}{8.4}{\rmdefault}{\mddefault}{\updefault}{\color[rgb]{0,0,0}$\theta_{100} = \{Z_1/\underline{X_{100}}\}$}%
}}}
\put(1351,-2011){\makebox(0,0)[lb]{\smash{\SetFigFont{7}{8.4}{\rmdefault}{\mddefault}{\updefault}{\color[rgb]{0,0,0}$C_{p_2}$}%
}}}
\put(1276,-2311){\makebox(0,0)[lb]{\smash{\SetFigFont{9}{10.8}{\rmdefault}{\mddefault}{\updefault}{\color[rgb]{0,0,0}$N_{101}$:  $q$ }%
}}}
\put(1276,-3361){\makebox(0,0)[lb]{\smash{\SetFigFont{9}{10.8}{\rmdefault}{\mddefault}{\updefault}{\color[rgb]{0,0,0}$N_{103}$:  $q$ }%
}}}
\put(1276,-2836){\makebox(0,0)[lb]{\smash{\SetFigFont{9}{10.8}{\rmdefault}{\mddefault}{\updefault}{\color[rgb]{0,0,0}$N_{102}$:  $q$ }%
}}}
\put(1351,-1186){\makebox(0,0)[lb]{\smash{\SetFigFont{7}{8.4}{\rmdefault}{\mddefault}{\updefault}{\color[rgb]{0,0,0}$C_{p_1}$}%
}}}
\put(1726,-661){\makebox(0,0)[lb]{\smash{\SetFigFont{7}{8.4}{\rmdefault}{\mddefault}{\updefault}{\color[rgb]{0,0,0}$\theta_1 = \{\underline{X_1}/f(X_2),Y_2/s(0)\}$}%
}}}
\put(1726,-136){\makebox(0,0)[lb]{\smash{\SetFigFont{7}{8.4}{\rmdefault}{\mddefault}{\updefault}{\color[rgb]{0,0,0}$\theta_0 = \{\underline{I}/f(X_1),Y_1/0\}$}%
}}}
\put(1351,-661){\makebox(0,0)[lb]{\smash{\SetFigFont{7}{8.4}{\rmdefault}{\mddefault}{\updefault}{\color[rgb]{0,0,0}$C_{p_1}$}%
}}}
\put(1351,-136){\makebox(0,0)[lb]{\smash{\SetFigFont{7}{8.4}{\rmdefault}{\mddefault}{\updefault}{\color[rgb]{0,0,0}$C_{p_1}$}%
}}}
\put(1276,-961){\makebox(0,0)[lb]{\smash{\SetFigFont{9}{10.8}{\rmdefault}{\mddefault}{\updefault}{\color[rgb]{0,0,0}$N_2$:  $p(\underline{X_2},s(s(0)))$ }%
}}}
\put(1276,-436){\makebox(0,0)[lb]{\smash{\SetFigFont{9}{10.8}{\rmdefault}{\mddefault}{\updefault}{\color[rgb]{0,0,0}$N_1$:  $p(\underline{X_1},s(0))$ }%
}}}
\put(1276,-1486){\makebox(0,0)[lb]{\smash{\SetFigFont{9}{10.8}{\rmdefault}{\mddefault}{\updefault}{\color[rgb]{0,0,0}$N_{100}$:  $p(\underline{X_{100}}, \underbrace{s(s(\ ...\ s}_{100 \ items}(0)\ ...\ )))$ }%
}}}
\put(1276, 89){\makebox(0,0)[lb]{\smash{\SetFigFont{9}{10.8}{\rmdefault}{\mddefault}{\updefault}{\color[rgb]{0,0,0}$N_0$:  $p(\underline{I}, 0)$}%
}}}
\thinlines
{\color[rgb]{0,0,0}\put(1651,-1681){\vector( 0,-1){480}}
}%
{\color[rgb]{0,0,0}\put(1651,-2911){\vector( 0,-1){300}}
}%
{\color[rgb]{0,0,0}\put(1651,-2386){\vector( 0,-1){300}}
}%
{\color[rgb]{0,0,0}\put(1651,-511){\vector( 0,-1){300}}
}%
{\color[rgb]{0,0,0}\put(1651, 14){\vector( 0,-1){300}}
}%
\end{picture}
\end{center}
\caption{A moded generalized SLDNF-tree generated 
by applying LP-check with Heuristic 1}\label{fig-eg5} 
\end{figure}
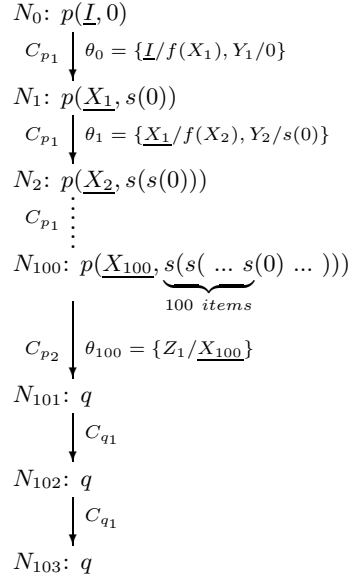

\begin{example}
\label{eg-p5}
{\em
Consider the logic program $P_5$ again. Let us choose $r = 3$.
By enhancing LP-check with Heuristic 1, 
Algorithm \ref{alg1} builds a moded generalized SLDNF-tree for the moded query
$Q_0 = p({\cal I}, 0)$ as shown in Figure \ref{fig-eg5}. Note that the first three nodes
satisfy conditions (a) and (b) of LP-check but violate condition (c). Although 
the second argument of $p(.)$ grows from $N_0$ through $N_1$ to $N_2$, it
has not grown to its maximum $p_{max}^2 = 100$. 
So the extension continues until it reaches $N_{100}$.
The three nodes $N_{98}$, $N_{99}$ and $N_{100}$ satisfy conditions (a),
(b) and (c). Since they do not satisfy Condition (c'), 
Algorithm \ref{alg1} cuts the derivation by skipping the clause $C_{p_1}$
for $N_{100}$. When the extension goes to $N_{103}$, the 
three nodes $N_{101}$, $N_{102}$ and $N_{103}$ satisfy conditions (a),
(b) and (c) and Condition (c'), thus Algorithm \ref{alg1} returns 
{\em most likely non-terminating} for $p({\cal I}, 0)$. 

As opposed to $Q_0$, for another interesting moded query 
$Q_1 = p({\cal I}, \underbrace{s(s(...s}_{101 \ items}(0)...)))$, 
Algorithm \ref{alg1} will yield an answer {\em most likely terminating}.     
}
\end{example} 

\subsection{Two Optimization Strategies}

Algorithm \ref{alg1} establishes a general framework 
for dynamic termination analysis of general logic programs 
with concrete or moded queries. It claims {\em terminating}/{\em non-terminating} 
only if the answer is provably terminating/non-terminating (see Theorem \ref{th-terminating}); 
otherwise it gives an approximate answer: 
{\em most likely terminating} or {\em most likely non-terminating}.
Although exploring all possible provably correct cases
is beyond the scope of this paper (an interesting topic
for further work), we identify the following two simple yet 
commonly occurring cases. 

For a logic program $P$ and a moded query $Q_0$,
assume that Algorithm \ref{alg1} 
encounters a partial derivation $D$ 
\begin{equation} 
\label{eq3-1}
N_0:G_0\Rightarrow_{C_0} ... N_{g_1}:G_{g_1}\Rightarrow_{C_k}   
...N_{g_2}:G_{g_2}\Rightarrow_{C_k} ... N_{g_r}:G_{g_r} 
\end{equation} 
that contains no negation arc ``$\cdot\cdot\cdot\triangleright$"
and satisfies the conditions of LP-check, where  
for any $j < r$, $L_{g_{j+1}}^1$ is a variant of $L_{g_j}^1$
and the sequence $S_{cl}$ of clauses applied between $N_{g_j}$ and $N_{g_{j+1}}$ 
is the same as the sequence between $N_{g_{j-1}}$ and $N_{g_j}$.

\begin{theorem}[Optimization Strategy 1]
$P$ is non-terminating for $Q_0$
if $D$ satisfies Condition (c') and $L = 0$. 
\end{theorem}

\noindent {\bf Proof:}
Since we use a fixed depth-first, left-most control strategy
and $D$ contains no negation arc \footnote{When a derivation contains a negation arc
like $N_i:\neg A \cdot\cdot\cdot\triangleright N_{i+1}:A$,
the evaluation of $A$ at $N_{i+1}$ will stop once one success
derivation for $A$ is generated. Some (infinite) derivations for $A$
may then be skipped.},
$D$ will be extended towards an infinite derivation $D'$ by
repeatedly applying the same sequence $S_{cl}$ of clauses, 
thus leading to an infinite number of loop goals 
$N_{g_{r+1}}:G_{g_{r+1}}$, $N_{g_{r+2}}:G_{g_{r+2}}$, ...,
where for each $i\geq 0$, $L_{g_{r+i+1}}^1$ is a variant of $L_{g_{r+i}}^1$.
Since $D$ satisfies Condition (c'), $D'$ also satisfies Condition (c')
because we apply the same sequence of clauses to variants of subgoals.
By Lemma \ref{lem1}, $D'$ must have an infinite moded instance. 
The condition $L = 0$ indicates that Algorithm \ref{alg1} 
never incorrectly cuts any derivations before, 
hence $P$ is non-terminating for $Q_0$. $\Box$ 

\begin{theorem}[Optimization Strategy 2]
If $D$ does not satisfy Condition (c') and 
for each $j \leq r$, $G_{g_j}$ contains only one subgoal, 
then the moded generalized SLDNF-tree $GT_{G_0}$ contains an infinite derivation  
with an infinite moded instance if and only if it contains
an infinite derivation with an infinite moded instance after 
skipping the clause $C_k$ at $N_{g_r}$. 
\end{theorem}

\noindent {\bf Proof:}
Following the above proof of Optimization Strategy 1,
when $D$ does not satisfy Condition (c'), $D'$ does not 
satisfy Condition (c'), either. By Lemma \ref{lem1}, 
$D'$ has no infinite moded instance. 

For simplicity, let $S_{cl}$ be the sequence of two clauses,
$C_k, C_{k'}$. Assume for each $j\geq 1$, we have derivation steps
\[N_{g_j}:G_{g_j}\Rightarrow_{C_k} N_{g_j'}:G_{g_j'} \Rightarrow_{C_{k'}} N_{g_{j+1}}:G_{g_{j+1}}\]
Since $L_{g_{j+1}}^1$ is a variant of $L_{g_j}^1$,
$L_{g_{j+1}'}^1$ is a variant of $L_{g_j'}^1$.

Let us cut $D'$ at 
$N_{g_r}$ (i.e. extend $D$ with the clause $C_k$ skipped).
Assume that $GT_{G_0}$ has an infinite derivation $D''$ 
with an infinite moded instance before this cut, and that 
on the contrary it has no infinite derivation 
with an infinite moded instance after the cut. 
$D''$ must be an extension of $D$ by repeatedly applying 
$C_k, C_{k'}$ for a certain number of times and then
at some node $N_{g_{r+m}}:G_{g_{r+m}}$ (or $N_{g_{r+m}'}:G_{g_{r+m}'}$),
skipping $C_k$ (or $C_{k'}$) to go towards an infinite derivation.
Since $L_{g_{r+m}}^1$ is a variant of $L_{g_{r-1}}^1$,
(resp. $L_{g_{r+m}'}^1$ is a variant of $L_{g_{r-1}'}^1$),
a copy (up to variable renaming) of the infinite 
derivation starting from $N_{g_{r+m}}:G_{g_{r+m}}$ 
(or from $N_{g_{r+m}'}:G_{g_{r+m}'}$) will appear starting from $N_{g_{r-1}}$
(or from $N_{g_{r-1}'}$). This copy of an infinite derivation
has the same infinite moded instances as $D''$.
This contradicts our assumption.

The above proof shows that if $GT_{G_0}$ has an infinite derivation  
with an infinite moded instance, then it has an infinite derivation 
with an infinite moded instance after skipping $C_k$ at $N_{g_r}$.
Since $D$ is negation-free (with no negation arcs), the converse also holds.
This proves the correctness of Optimization Strategy 2. $\Box$

\bigskip 

When the condition of Optimization Strategy 2 holds,
we can extend $D$ with the clause $C_k$ skipped safely.
Therefore, in this case we do not need to set $L = 1$
in Algorithm \ref{alg1} (setting $L = 1$ leads to an approximate answer).

Plugging the above two strategies into Algorithm \ref{alg1} gives rise to the following new algorithm.
\begin{algorithm} 
\label{alg2} 
{\em 
Testing termination of a logic program $P$ for a moded query $Q_0$,
given a repetition number $r\geq 3$. 
\begin{enumerate} 
\item
Initially, set $L = 0$. 
Construct the moded generalized SLDNF-tree $GT_{G_0}$, where for each 
partial derivation $D$ satisfying the conditions of LP-check,
if $D$ satisfies Condition (c') then goto \ref{no}, else set $L = 1$
unless the condition of Optimization Strategy 2 holds, 
and extend $D$ with clause $C_k$ skipped.

\item
Return {\em terminating} if $L = 0$; otherwise return {\em most likely terminating}  

\item
Return {\em non-terminating} if the condition of 
Optimization Strategy 1 holds; otherwise return {\em most likely non-terminating}. 
\end{enumerate} 
} 
\end{algorithm}  

\begin{example}
\label{eg-more}
{\em
We test the termination of logic programs, $P_0 - P_5$, again by
applying Algorithm \ref{alg2}.

For $P_0$, the partial derivation (Figure \ref{fig1}) between $N_0$ and $N_4$
satisfies the condition of Optimization Strategy 2,
so Algorithm \ref{alg2} skips $C_{p_2}$. It keeps $L = 0$ till the end and
returns {\em terminating} for $Q_0$ (Algorithm \ref{alg1} 
returns {\em most likely terminating}). 

For $P_1$, neither of the two strategies is
applicable, so Algorithm \ref{alg2} returns {\em most likely non-terminating}
for $p({\cal I})$ as Algorithm \ref{alg1} does.

For $P_2$, Algorithm \ref{alg2} returns {\em terminating} for $p({\cal I})$.
For a query $q$, its derivation satisfies the condition of Optimization Strategy 1,
thus leading to an answer {\em non-terminating} (Algorithm \ref{alg1} 
returns {\em most likely non-terminating}). 
 
For $P_3$, like Algorithm \ref{alg1}, Algorithm \ref{alg2}   
builds $GT_{G_0^1}$, $GT_{G_0^2}$ and $GT_{G_0^3}$
as shown in Figures \ref{fig2} (a), (b) and (c), respectively
for the three moded queries. 
The derivations in $GT_{G_0^1}$ and $GT_{G_0^3}$
satisfy the condition of Optimization Strategy 2, so
Algorithm \ref{alg2} keeps $L = 0$ till the end and
returns an answer {\em terminating}
(Algorithm \ref{alg1} returns {\em most likely terminating})
for $Q_0^1$ and $Q_0^3$. 
The derivation in $GT_{G_0^2}$ satisfies
the condition of Optimization Strategy 1, so
Algorithm \ref{alg2} returns an answer {\em non-terminating}
(Algorithm \ref{alg1} returns {\em most likely non-terminating}) for $Q_0^2$.
By Theorem \ref{th-inc}, we infer that $P_3$ terminates 
for all moded queries in $MQ(P_3)$ except for $Q_0^2$
that does not terminate.
 
For $P_4$, neither of the two strategies is
applicable, so Algorithm \ref{alg2} returns the same 
answers as Algorithm \ref{alg1}.

For $P_5$, Algorithm \ref{alg2} returns the same 
answers as Algorithm \ref{alg1} for any moded queries
with a predicate symbol $p$. For the query $q$, Optimization Strategy 1 
applies, so Algorithm \ref{alg2} returns an answer {\em non-terminating}.  
}
\end{example} 

\section{Related Work}
\label{related-work}
Termination of a logic program can be addressed in either a static
or a dynamic way. Static termination analysis builds from 
the source code of a logic program some well-founded 
termination conditions/constraints in terms of level mappings,
interargument size relations and/or instantiation dependencies  
\cite{Apt1,Bezem92,BCF94,BS91,DV95,DDF93,DDV99,GC05,LS97,MN01,Pl90a,UVG88}.
In contrast, a dynamic termination approach characterizes and tests 
termination of a logic program by applying a loop checking technique.
It directly makes use of some essential dynamic characteristics of infinite derivations,
such as repetition of variant subgoals and recursive 
increase in term size, which are hard to capture in a static way
(for example, it is difficult to apply static termination analysis
to prove that $P_2$ terminates for a moded query $p({\cal I})$ and that
$P_5$ terminates for $p({\cal I}, \underbrace{s(s(...s}_{101 \ items}(0)...)))$
but does not terminate for $p({\cal I}, 0)$).
This paper develops a new dynamic approach with a characterization and a
testing algorithm for moded queries. To the best of 
our knowledge, no similar work has been reported in the literature.

The core of a dynamic termination approach is a characterization of infinite
derivations. In \cite{shen-tocl}, the first such characterization
is established for general logic programs. However, it applies only
to concrete queries and cannot handle moded queries.
 
A dynamic termination approach uses a loop checking mechanism (a loop check)
to implement a characterization of infinite derivations. 
Representative loop checks include VA-check \cite{BAK91,VG87}, 
OS-check \cite{BDM92,MD96,Sa93}, and VAF-checks \cite{shen97,shen001}.
All apply to positive logic programs. In particular,
VA-check applies to function-free logic programs, where an 
infinite derivation is characterized by a sequence of selected {\em variant subgoals}. 
OS-check identifies an infinite derivation with a sequence of selected subgoals 
with the same predicate symbol {\em whose sizes do not decrease}.
VAF-checks take a sequence of selected {\em expanded variant subgoals} 
as major characteristics of an infinite derivation. Expanded variant subgoals are variant
subgoals except that some terms may grow bigger. In this paper, 
a new loop check mechanism, LP-check (with Heuristic 1), is introduced in which 
an infinite derivation is identified with a sequence of {\em loop goals}.
LP-check is more effective than VA-check, OS-check and VAF-checks,
none of which can handle the logic program $P_5$. Most importantly, enhancing
LP-check with Condition (c') leads to the first loop check for moded queries.

\section{Conclusion and Future Work}
\label{conclusion}
We have presented a dynamic approach to characterizing and testing termination of
a general logic program. The approach is very powerful and useful. It can be used 
(1) to test if a logic program terminates for a given concrete or moded query,
(2) to test if a logic program terminates for all concrete or moded queries, and
(3) to find all (most general) concrete/moded queries that are most likely  
terminating (or non-terminating). For any concrete/moded query, the algorithm 
yields an answer {\em terminating}, 
{\em most likely terminating}, {\em non-terminating} or {\em most likely non-terminating}.
For a great majority of representative logic programs we collected in the literature,  
an answer {\em most likely terminating} (resp. {\em most likely non-terminating})
implies terminating (resp. non-terminating).
The algorithm can be incorporated into Prolog as a debugging tool,
which would provide the user with valuable debugging 
information for him/her to understand the causes of non-termination.   
 
A conspicuous advantage of a dynamic termination approach over static
termination analysis is that it tests termination on the fly (i.e. by evaluating 
some queries), thus capturing essential characteristics of infinite
derivations. This makes a dynamic approach able to guess if a partial
derivation is most likely to extend towards an infinite one.
Although static termination analysis has been extensively studied
over the last few decades, exploration of dynamic termination approaches is just at
the beginning. We expect to see more prosperous research in this direction. 
Many problems are open, including extensions to typed queries \cite{BCGGV05}
and to logic programs with tabling \cite{chen96,shen004,VDK001}.
Our ongoing work aims to develop a dynamic termination analyser 
and make a comparative study with existing static termination analysers.
It is also promising future work to combine static and dynamic approaches
for a hybrid termination analyser. 

\end{document}